\documentclass[11pt]{article}
\pdfoutput=1
\usepackage[T1]{fontenc}
\usepackage[utf8]{inputenc}
\usepackage{amsmath,amssymb}
\usepackage{mathptmx}
\usepackage{textcomp}
\usepackage{newunicodechar}
\newunicodechar{ρ}{\ensuremath{\rho}}
\newunicodechar{β}{\ensuremath{\beta}}
\newunicodechar{τ}{\ensuremath{\tau}}
\newunicodechar{σ}{\ensuremath{\sigma}}
\newunicodechar{α}{\ensuremath{\alpha}}
\newunicodechar{γ}{\ensuremath{\gamma}}
\newunicodechar{δ}{\ensuremath{\delta}}
\newunicodechar{ε}{\ensuremath{\varepsilon}}
\newunicodechar{μ}{\ensuremath{\mu}}
\newunicodechar{π}{\ensuremath{\pi}}
\newunicodechar{φ}{\ensuremath{\phi}}
\newunicodechar{χ}{\ensuremath{\chi}}
\newunicodechar{θ}{\ensuremath{\theta}}
\newunicodechar{λ}{\ensuremath{\lambda}}
\newunicodechar{ν}{\ensuremath{\nu}}
\newunicodechar{η}{\ensuremath{\eta}}
\newunicodechar{κ}{\ensuremath{\kappa}}
\newunicodechar{ω}{\ensuremath{\omega}}
\newunicodechar{ψ}{\ensuremath{\psi}}
\newunicodechar{ξ}{\ensuremath{\xi}}
\newunicodechar{Δ}{\ensuremath{\Delta}}
\newunicodechar{Φ}{\ensuremath{\Phi}}
\newunicodechar{Σ}{\ensuremath{\Sigma}}
\newunicodechar{Ω}{\ensuremath{\Omega}}
\newunicodechar{Γ}{\ensuremath{\Gamma}}
\newunicodechar{Λ}{\ensuremath{\Lambda}}
\newunicodechar{Ψ}{\ensuremath{\Psi}}
\newunicodechar{−}{\ensuremath{-}}
\newunicodechar{×}{\ensuremath{\times}}
\newunicodechar{≥}{\ensuremath{\geq}}
\newunicodechar{≤}{\ensuremath{\leq}}
\newunicodechar{≈}{\ensuremath{\approx}}
\newunicodechar{≠}{\ensuremath{\neq}}
\newunicodechar{±}{\ensuremath{\pm}}
\newunicodechar{√}{\ensuremath{\surd}}
\newunicodechar{∞}{\ensuremath{\infty}}
\newunicodechar{→}{\ensuremath{\rightarrow}}
\newunicodechar{←}{\ensuremath{\leftarrow}}
\newunicodechar{↔}{\ensuremath{\leftrightarrow}}
\newunicodechar{⇒}{\ensuremath{\Rightarrow}}
\newunicodechar{∑}{\ensuremath{\sum}}
\newunicodechar{∈}{\ensuremath{\in}}
\newunicodechar{∼}{\ensuremath{\sim}}
\newunicodechar{∝}{\ensuremath{\propto}}
\newunicodechar{′}{\ensuremath{'}}
\newunicodechar{·}{\ensuremath{\cdot}}
\newunicodechar{⋯}{\ensuremath{\cdots}}
\newunicodechar{…}{\ldots{}}
\newunicodechar{∣}{\ensuremath{\mid}}
\newunicodechar{≡}{\ensuremath{\equiv}}
\newunicodechar{⊥}{\ensuremath{\perp}}
\newunicodechar{∂}{\ensuremath{\partial}}
\newunicodechar{∫}{\ensuremath{\int}}
\newunicodechar{²}{\textsuperscript{2}}
\newunicodechar{³}{\textsuperscript{3}}
\newunicodechar{¹}{\textsuperscript{1}}
\newunicodechar{⁴}{\textsuperscript{4}}
\newunicodechar{₀}{\textsubscript{0}}
\newunicodechar{₁}{\textsubscript{1}}
\newunicodechar{₂}{\textsubscript{2}}
\newunicodechar{ᵏ}{\textsuperscript{k}}
\newunicodechar{†}{\textdagger{}}
\newunicodechar{‡}{\textdaggerdbl{}}
\newunicodechar{§}{\S{}}
\newunicodechar{½}{\ensuremath{\tfrac12}}
\newunicodechar{°}{\textdegree{}}
\newunicodechar{✓}{\checkmark{}}
\newunicodechar{‑}{-}
\newunicodechar{ˆ}{\^{}}
\newunicodechar{≳}{\ensuremath{\gtrsim}}
\newunicodechar{≲}{\ensuremath{\lesssim}}
\usepackage[letterpaper,margin=1in]{geometry}
\usepackage{microtype}
\usepackage{graphicx}
\usepackage{longtable,booktabs,array,calc}
\usepackage{placeins}
\usepackage{etoolbox}
\makeatletter
\patchcmd\longtable{\par}{\if@noskipsec\mbox{}\fi\par}{}{}
\def\maxwidth{\ifdim\Gin@nat@width>\linewidth\linewidth\else\Gin@nat@width\fi}
\def\maxheight{\ifdim\Gin@nat@height>\textheight\textheight\else\Gin@nat@height\fi}
\def\fps@figure{tbp}
\makeatother
\setkeys{Gin}{width=\maxwidth,height=\maxheight,keepaspectratio}
\usepackage[labelsep=period,labelfont=bf,font=small]{caption}
\usepackage{authblk}

\usepackage[numbers,sort&compress]{natbib}
\setcitestyle{square}

\usepackage{xcolor}
\usepackage[hidelinks]{hyperref}
\usepackage{xurl}
\hypersetup{pdftitle={Randomization protects the average, not the trial},pdfauthor={William J. Dwyer; David Zurakowski}}

\title{Randomization protects the average, not the trial: chance covariate imbalance and a calibration of balance \emph{P} values at the design stage}
\author[1]{William J. Dwyer\thanks{Corresponding author. Southwick Hall 303, One University Ave, Lowell, MA 01854, USA. Email: \href{mailto:William_Dwyer@student.uml.edu}{William\_Dwyer@student.uml.edu}. ORCID \href{https://orcid.org/0009-0004-0855-7222}{0009-0004-0855-7222}.}}
\author[2,3]{David Zurakowski\thanks{ORCID \href{https://orcid.org/0000-0003-3610-6942}{0000-0003-3610-6942}.}}
\affil[1]{Department of Mathematics and Statistics, University of Massachusetts Lowell, Lowell, MA, USA}
\affil[2]{Department of Anesthesiology, Critical Care and Pain Medicine and Department of Surgery, Boston Children's Hospital, Boston, MA, USA}
\affil[3]{Harvard Medical School, Boston, MA, USA}
\date{25 November 2026}
\begin{document}
\maketitle
\begingroup\renewcommand\thefootnote{}\footnotetext{Preprint, version 3 (25 November 2026). Reproducibility code: Zenodo, version 1.3, \href{https://doi.org/10.5281/zenodo.24520628}{doi:10.5281/zenodo.24520628} (all versions: \href{https://doi.org/10.5281/zenodo.24210820}{doi:10.5281/zenodo.24210820}). Licence: CC BY 4.0.}\endgroup
\begin{abstract}
\textbf{Background.} Randomization makes the unadjusted treatment-effect estimate unbiased across the allocations a scheme could produce, but chance imbalance remains possible in a single trial. Design-stage tools score randomization schemes by simulated balance \emph{P} values, but their implications for the analysis are seldom stated. We review chance imbalance and derive that link.

\textbf{Methods.} A targeted narrative review of methodological, simulation and re-analysis studies, regulatory guidance and CONSORT 2025, with analytic derivations under a linear model with one normally distributed prognostic covariate, and a simulation comparing six randomization schemes. Derivations were checked by simulation; code is supplied.

\textbf{Results.} Given the realized allocation, the unadjusted estimate is biased by β × (difference in covariate means). In standard-error units it equals ρ × Z\textsubscript{bal} (ρ, the covariate--outcome correlation; Z\textsubscript{bal}, the balance z-statistic), so it does not shrink as trials grow. Correctly specified ANCOVA removes this bias. Change scores can reverse it. Marginal and conditional odds or hazard ratios can differ under perfect balance. The Sellke--Bayarri--Berger bound reaches 1 at 1/e. For one balance \emph{P} value, 1/e maximizes first-order flag-rate separation under a Beta shift. Under complete randomization the worst of k independent balance \emph{P} values falls below 1/e with probability 1 − (1 − 1/e)\textsuperscript{k} (0.99 for k = 10). For one covariate, any balance \emph{P} value ≥ 2Φ(−1) = 0.3173\ldots{} keeps the two-sided conditional type I error of the unadjusted test at most 5\% for every ρ; at 1/e the one-sided error in the favoured direction is at most about 4\%, against 13--33\% at \emph{P} = 0.05 for ρ of 0.5--0.9. In simulation, permuted blocks and Efron's coin scored similarly to complete randomization. Stratification and minimization improved balance at greater predictability. Pre-specified ANCOVA stayed near nominal under all six schemes.

\textbf{Conclusions.} Compare scheme scores against complete randomization on the same covariates. Report prognostic strength, predictability and an analysis plan adjusting for every balancing factor. For one covariate, 1/e has a mathematical calibration. With several covariates, imbalances combine by sign and prognostic weight, so per-covariate thresholds do not bound the total. Clinical validation remains open.
\end{abstract}


\hypertarget{background}{%
\section{Background}\label{background}}

Randomization makes the unadjusted estimator of the treatment effect unbiased over the randomization distribution, protects against selection bias when allocation is concealed, and supplies \emph{P} values and confidence intervals whose error rates already allow for chance imbalance~\cite{senn2013,rosenberger2008}. It does not balance the trial that is actually run. Remaining baseline differences ``are the result of chance rather than bias''~\cite{consort2025ee}, and in any single trial they can favour one arm if the analysis ignores them~\cite{kahan2014}.

The question of how much this matters now has a practical form. Software that simulates thousands of allocations under candidate schemes, and scores each scheme by the distribution of baseline balance statistics, is used to choose among complete randomization, permuted and stratified blocks, urn designs and minimization at the design stage; one of us first described this approach in a 2013 patent application~\cite{dwyer2013}. Such scores are usually expressed as balance \emph{P} values with a cut-off that flags imbalance. The debate over baseline significance testing~\cite{altman1985,altmandore1990,permutt1990,begg1990,senn1994,mutz2019,consort2025ee} and the literature that quantifies imbalance bias~\cite{senn1989,egbewale2014,kahan2014,ciolino2015} have rarely been connected to this use. Senn showed that passing a baseline test does not control the conditional type I error~\cite{senn1989,senn1993,senn1994}. Ciolino and colleagues tabulated the balance t statistics at which an unadjusted analysis loses power or doubles its type I error, and suggested that, for judging a realized allocation, a one-sided baseline significance level as high as about 24\% would be more appropriate than 5\%~\cite{ciolino2015}. Zhao and Ding gave a randomization-based theory for rerandomization that accepts an allocation only when covariate balance \emph{P} values clear a threshold~\cite{zhaoding2024rep}. We found no published account of how to choose and justify a threshold for scoring and comparing randomization schemes at the design stage.

This article has three aims: to summarize what is known about chance covariate imbalance and how the analysis carries or removes it, to derive four properties that govern how a design-stage balance score built on \emph{P} values should be read, and to compare common randomization schemes on such a score by simulation, with items to report alongside it. Material that most trialists know, and the reference tables behind the figures, are in Additional file 1.

\hypertarget{methods}{%
\section{Methods}\label{methods}}

\hypertarget{literature-review}{%
\subsection{Literature review}\label{literature-review}}

We conducted a targeted narrative review rather than a systematic review, and report it with reference to the SANRA criteria for narrative reviews~\cite{baethge2019}. Sources were identified between October and November 2026 by searches of the web, PubMed Central, arXiv and regulatory agency websites, using combinations of the terms covariate imbalance, baseline imbalance, chance bias, accidental bias, conditional bias, ANCOVA, change score, non-collapsibility, stratified randomization, minimization, rerandomization, baseline significance testing, balance test and conditional size, and by following citations from key papers. We included methodological papers, simulation studies, re-analyses of individual trials, meta-epidemiological studies, regulatory and reporting guidance, and widely cited commentaries. Every cited source was read in full text, in its published version. Every direct quotation and specific figure attributed to a source was checked against that text. Regulatory documents were taken from the issuing agencies: the FDA guidance on covariate adjustment (May 2023)~\cite{fda2023}, the EMA guideline on adjustment for baseline covariates (2015)~\cite{ema2015}, ICH E9~\cite{iche9} and CONSORT 2025~\cite{consort2025ee}.

\hypertarget{analytic-framework}{%
\subsection{Analytic framework}\label{analytic-framework}}

We consider a two-arm trial with n patients per arm, allocated by complete randomization with equal arm sizes, and the linear outcome model

\[Y = \mu + \tau T + \beta X + \varepsilon, \qquad (1)\]

where T = 1 for treatment and 0 for control, X is a baseline covariate with variance σ\textsubscript{X}\textsuperscript{2}, and ε is independent of X and T with variance σ\textsubscript{ε}\textsuperscript{2}. The outcome variance is σ\textsubscript{Y}\textsuperscript{2} = β\textsuperscript{2}σ\textsubscript{X}\textsuperscript{2} + σ\textsubscript{ε}\textsuperscript{2} and the covariate--outcome correlation is ρ = βσ\textsubscript{X}/σ\textsubscript{Y}. Let Δ\textsubscript{X} = \ensuremath{\bar{x}}\textsubscript{T} − \ensuremath{\bar{x}}\textsubscript{C}, d = Δ\textsubscript{X}/σ\textsubscript{X} the standardized mean difference, and \ensuremath{\hat{\tau}} the unadjusted difference in mean outcomes. The balance statistic is

\[Z_{\text{bal}} = \frac{\Delta_X}{\sigma_X\sqrt{2/n}}, \qquad (2)\]

and the unadjusted test statistic for H\textsubscript{0}: τ = 0 is T\textsubscript{u} = \ensuremath{\hat{\tau}}/(σ\textsubscript{Y}√(2/n)). Variances are treated as known, which is exact for normal data with known variances and holds asymptotically otherwise. For binary outcomes we use a binary prognostic factor H with stratum risks r\textsubscript{H} and r\textsubscript{L} and a treatment effect constant on the risk-difference scale. Throughout, the balance \emph{P} value for one covariate is \emph{P} = 2\{1 − Φ(\textbar Z\textsubscript{bal}\textbar)\}, the \emph{worst balance P value} of a trial is the smallest \emph{P} value over the covariates scored, and Pr(·) denotes probability.

For calibration we use the bound of Sellke, Bayarri and Berger~\cite{sellke2001}, B(\emph{P}) = −e·\emph{P}·ln(\emph{P}) for \emph{P} \textless{} 1/e and B(\emph{P}) = 1 otherwise, a lower bound on the Bayes factor of the null to the alternative over a broad class of alternatives, with associated conditional error probability α(\emph{P}) = \{1 + 1/B(\emph{P})\}\textsuperscript{−1}.

\hypertarget{computation}{%
\subsection{Computation}\label{computation}}

Derivations are given in full in Additional file 1. All results marked as derived were computed in Python 3 with NumPy and SciPy~\cite{virtanen2020}, and the analytic results for conditional bias, change-score bias, ANCOVA precision, conditional type I error and the binary-outcome bias were each checked against Monte Carlo simulation, including a check with estimated variances and t-tests at 10 and 25 patients per arm (Additional file 2). The comparison of randomization schemes was simulated with the same tools (Additional file 4). Figures were drawn with Matplotlib~\cite{hunter2007} directly from the same formulas and from fixed-seed simulations (Additional files 3 and 5). Statements in the text that rest on these derivations are marked \emph{derived}; they are standard algebra or new calculations, not quotations from the cited papers.

\hypertarget{results}{%
\section{Results}\label{results}}

\hypertarget{randomization-protects-the-average-not-the-trial}{%
\subsection{Randomization protects the average, not the trial}\label{randomization-protects-the-average-not-the-trial}}

Over all allocations a scheme can produce, E\textsubscript{A}{[}\ensuremath{\hat{\tau}}{]} = τ. Given the realized allocation A, under model (1),

\[E[\hat\tau \mid A] = \tau + \beta\,\Delta_X. \qquad (3)\]

The second term averages to zero over allocations, which is why unconditional inference remains valid, but it is not zero in a trial in which a prognostic covariate is imbalanced (\emph{derived}; D1; Additional file 1, Table A1). Senn's resolution is that ``randomisation entitles you to ignore unmeasured covariates but that you must condition on what you observe''~\cite{senn2013}; in practice, conditioning means pre-specified adjustment. Restricted randomization keeps the average guarantee when arm sizes are equal and the balance criterion treats the arms symmetrically~\cite{morgan2012}; a scheme can change only the spread of conditional bias across the allocations it produces.

In standardized units the conditional bias is

\[\frac{\text{bias}}{\sigma_Y} = \rho\, d. \qquad (4)\]

A correlation of 0.5 and an imbalance of 0.2 SD give a bias of 0.1 SD, half of a conventional small effect (\emph{derived}; D2). When the baseline and the outcome are the same measurement, the analysis decides the sign. Final-value analysis carries the bias βΔ\textsubscript{X}; the change score Y − X has conditional bias (β − 1)Δ\textsubscript{X}, of opposite sign when 0 \textless{} β \textless{} 1; and ANCOVA estimates β and removes the term (\emph{derived}; D3)~\cite{egbewale2014,vickers2001}. Figure 1 shows all three in the same simulated trials: every analysis is unbiased on average, but only ANCOVA is unbiased given the allocation.

\begin{figure}
\centering
\includegraphics[width=1\textwidth,height=\textheight]{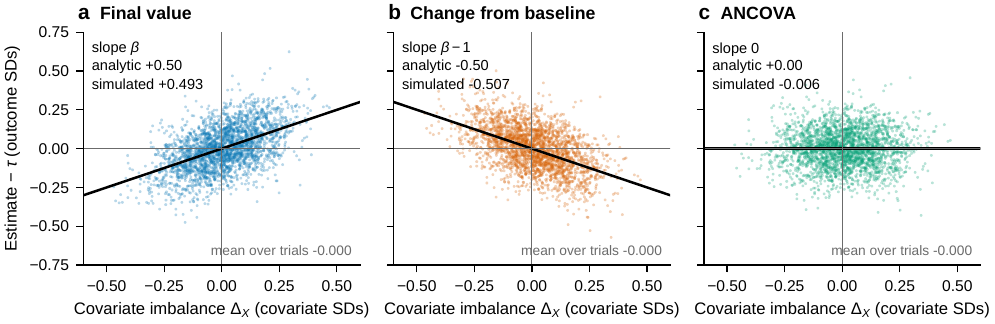}
\caption{\textbf{Conditional bias of three analyses of the same simulated trials.} Each point is one of 20 000 simulated two-arm trials with 100 patients per arm, complete randomization and no treatment effect; covariate and outcome are standardized and ρ = β = 0.5 (2500 trials plotted). The horizontal axis is the realized covariate imbalance Δ\textsubscript{X}; the vertical axis is the estimate minus the true effect. Black lines show the analytic conditional bias: (\textbf{a}) final-value analysis, βΔ\textsubscript{X} (equation 3); (\textbf{b}) change-from-baseline analysis, (β − 1)Δ\textsubscript{X}, of opposite sign; (\textbf{c}) ANCOVA, zero. Fitted slopes were 0.493, −0.507 and −0.006. All three analyses average to the true effect over trials (unbiased on average), but only ANCOVA is unbiased given the allocation drawn. Additional file 1, D1 and D3.}
\end{figure}

For a binary outcome with high- and low-risk strata and a treatment effect constant on the risk-difference scale, the conditional bias of the crude risk difference is

\[\text{bias} = (\pi_{H,T} - \pi_{H,C})(r_H - r_L), \qquad (5)\]

where π\textsubscript{H,T} and π\textsubscript{H,C} are the proportions of high-risk patients in each arm (\emph{derived}; D4). Ten percentage points more high-risk patients in the treated arm, with r\textsubscript{H} − r\textsubscript{L} = 0.30, give a bias of 3 percentage points against treatment, comparable with many absolute risk reductions that trials are powered to detect. In the NICHD Neonatal Research Network, 2013--2018, survival to discharge or 1 year among actively treated infants was 55.8\% at 23 weeks and 94.7\% at 28 weeks~\cite{bell2022}; a 10-point excess of 23-week infants in one arm gives a bias of about 3.9 percentage points in the mortality difference (\emph{derived}; arithmetic on published rates, not a published estimate). Late-onset sepsis is similarly prognostic, with mortality of 17--18\% in infected against 7\% in uninfected infants~\cite{stoll1996,stoll2002} (Figure 2).

\begin{figure}
\centering
\includegraphics[width=0.62\textwidth,height=\textheight]{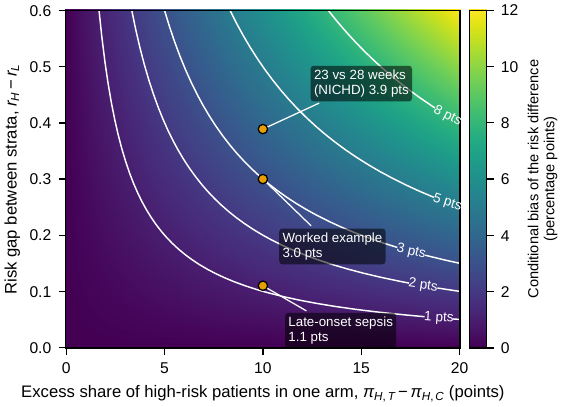}
\caption{\textbf{Conditional bias of the crude risk difference for a binary outcome.} Colour and white contours give the bias (π\textsubscript{H,T} − π\textsubscript{H,C})(r\textsubscript{H} − r\textsubscript{L}) of equation (5), in percentage points, as a function of the excess share of high-risk patients in one arm and the gap in event risk between high- and low-risk strata. Points mark the worked example (10 points of excess share, risk gap 0.30: 3.0 points), the gap in mortality between infants born at 23 and 28 weeks in the NICHD Neonatal Research Network (survival of actively treated infants 55.8\% and 94.7\%, gap 0.389: 3.9 points)~\cite{bell2022} and late-onset sepsis (mortality 18\% against 7\%, gap 0.11: 1.1 points)~\cite{stoll2002}. Illustrative arithmetic assuming a constant risk difference across strata. Additional file 1, D4.}
\end{figure}

\hypertarget{chance-imbalance-is-common-and-does-not-shrink-relative-to-precision}{%
\subsection{Chance imbalance is common and does not shrink relative to precision}\label{chance-imbalance-is-common-and-does-not-shrink-relative-to-precision}}

If k independent covariates are each tested at α = 0.05, the probability that at least one gives \emph{P} \textless{} 0.05 is 1 − 0.95\textsuperscript{k} at any sample size; Morgan and Rubin note that ``for a modest 10 covariates and a 5\% significance level, this probability is 40\%''~\cite{morgan2012} (\emph{derived}; D5; Additional file 1, Table A3). In absolute terms imbalance does shrink with n, because the standardized difference has variance 2/n under complete randomization (Figure 3a; Table A4). Relative to the precision of the trial it does not. Dividing the bias by SE(\ensuremath{\hat{\tau}}) = σ\textsubscript{Y}√(2/n) gives

\[\frac{\text{bias}}{\text{SE}(\hat\tau)} = \rho\, Z_{\text{bal}}, \qquad (6)\]

and Z\textsubscript{bal} is standard normal for every n (\emph{derived}; D6). The conditional bias in standard errors therefore has the same distribution in a trial of 50 patients as in one of 5000; only ρ changes it (Figure 3b, c). Senn reached the same conclusion for the size of the test of efficacy in 1989: covariate imbalance ``is just as much a problem for large studies as for small ones in terms of effect on size''~\cite{senn1989,senn1994}. Later, discussing the myth that ``large trials are more balanced than small ones'', he grants that larger trials are better balanced ``in proportionate terms'' but notes that ``the effect of increasing sample size is consumed by conventional analyses in terms of increased precision'', so that ``smaller differences at baseline become important''~\cite{senn2013}.

\begin{figure}
\centering
\includegraphics[width=1\textwidth,height=\textheight]{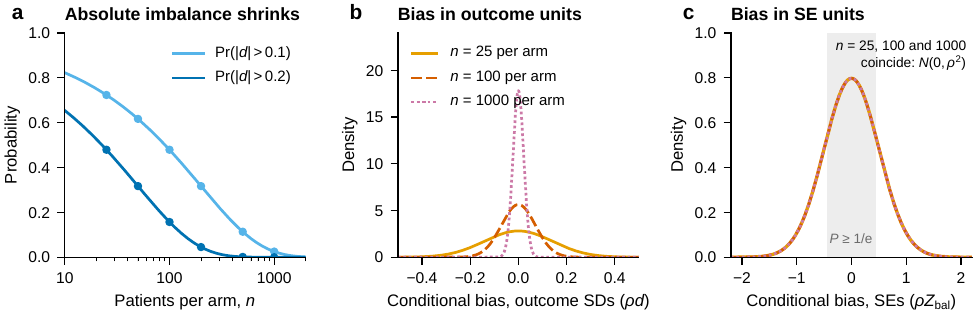}
\caption{\textbf{Chance imbalance shrinks with sample size in outcome units but not in standard-error units.} (\textbf{a}) Probability that the standardized covariate difference \textbar d\textbar{} exceeds 0.1 or 0.2 under complete randomization, against patients per arm; points are the entries of Additional file 1, Table A4. (\textbf{b}) Distribution of the conditional bias of the unadjusted estimate in outcome standard deviations, ρd \ensuremath{\sim} N(0, 2ρ\textsuperscript{2}/n), for ρ = 0.5 and 25, 100 and 1000 patients per arm. (\textbf{c}) The same bias in standard errors of the estimate, ρZ\textsubscript{bal} \ensuremath{\sim} N(0, ρ\textsuperscript{2}): the three curves coincide. The shaded band contains the bias when the balance \emph{P} value is at least 1/e. Additional file 1, D5 and D6.}
\end{figure}

Prognostic strength is what turns imbalance into harm. In an interim-analysis simulation with a true treatment effect, the probability of wrongly stopping for futility at an interim look at 30\% of the information was 0.22 at every imbalance level when ρ = 0, but rose to 0.58, 0.75 and 0.86 at ρ = 0.8 as imbalance grew~\cite{ciolino2014}. Kernan and colleagues concluded that stratification ``may prevent type I error and improve power for small trials (\textless{} 400 patients),'' but only when stratification factors strongly affect prognosis~\cite{kernan1999}.

\hypertarget{the-analysis-carries-reverses-or-removes-the-bias}{%
\subsection{The analysis carries, reverses or removes the bias}\label{the-analysis-carries-reverses-or-removes-the-bias}}

Across 126 simulated scenarios crossing effect size, baseline--outcome correlation and imbalance, the final-value analysis was biased in the direction of the imbalance, change-score bias ran the other way and was largest at low correlation, and ANCOVA remained unbiased with a standard error relative to ANOVA close to √(1 − ρ\textsuperscript{2}); the authors concluded that ``ANCOVA should be the analysis of choice, a priori''~\cite{egbewale2014}. Analysing change ``does not control for baseline imbalance because of regression to the mean''~\cite{vickers2001}. Overall and Magee found the same pattern by simulation: when baseline means differed by chance, tests on post-treatment scores alone were nonconservative in the direction of the baseline difference and tests on simple difference scores in the opposite direction, whereas ANCOVA was free of directional bias~\cite{overall1992}. In a re-analysis of 12 outcomes from 8 trials with unadjusted power fixed at 80\%, adjustment for known prognostic covariates raised median power to 92.6\%, while the worst loss from adjusting for a non-prognostic covariate was 1.5 points~\cite{kahan2014}; for binary outcomes, simulated sample-size reductions from adjustment ranged from 3\% to 46\% as the covariate effect grew, at 50\% outcome incidence and covariate prevalence, and were smaller at lower incidence or prevalence~\cite{hernandez2004}.

Ignoring the factors used to balance the allocation makes the analysis conservative. In simulations, unadjusted standard errors were 2.3\% to 17.4\% too large and type I error fell to 2.2--4.3\% against a nominal 5\%~\cite{kahan2012sim}; the FDA states that ``an analysis ignoring stratified randomization is likely to overestimate standard errors and can be unduly conservative''~\cite{fda2023,morris2022}. Yet among 258 trials published in 2010 in four general medical journals, 63\% balanced on at least one variable and only 26\% adjusted for all balancing factors~\cite{kahan2012bmj}. Post hoc adjustment chosen because of an observed imbalance is, in EMA's words, not ``an appropriate reason for including'' a covariate in the primary analysis~\cite{ema2015}; it is a sensitivity analysis, not a rescue. These analysis choices are summarized in Additional file 1, Table A2.

\hypertarget{non-collapsibility-is-a-different-phenomenon}{%
\subsection{Non-collapsibility is a different phenomenon}\label{non-collapsibility-is-a-different-phenomenon}}

In logistic and Cox models, omitting a prognostic covariate attenuates the treatment estimate towards the null even when the covariate is perfectly balanced; under randomization the asymptotic bias from an omitted covariate is zero, in regular cases, only when the regression is linear or exponential~\cite{gail1984}. The FDA calls the effect ``distinct from confounding'' and able to ``occur despite randomization and large sample sizes''~\cite{fda2023}. With two equal, perfectly balanced strata, control risks of 0.10 and 0.60 and a conditional odds ratio of 8.0, the marginal odds ratio is 4.27 (\emph{derived}; D12; Additional file 1, Figure A10). The unadjusted odds ratio is not biased; it estimates a different, population-averaged estimand, which the FDA asks sponsors to choose in advance~\cite{fda2023}. Adjusting a logistic model therefore moves the estimate away from 1 and widens its standard error, yet never makes the test of no treatment effect less efficient~\cite{robinson1991}. An adjusted odds ratio that moves away from 1 thus says nothing about imbalance by itself, whereas one that moves towards 1 suggests an imbalance that had inflated the unadjusted effect, an interpretation rather than a published finding (Additional file 1, Table A9).

\hypertarget{documented-trials-show-double-digit-shifts-and-lasting-disputes}{%
\subsection{Documented trials show double-digit shifts and lasting disputes}\label{documented-trials-show-double-digit-shifts-and-lasting-disputes}}

In the MIST2 pleural-infection trial, baseline effusion size averaged 39 in the placebo arm and 47 in the tPA plus DNase arm. Adjustment moved the treatment effect from −12.3 to −7.6, a 38\% reduction, while its standard error fell from 4.3 to 2.8~\cite{kahan2014}. The NINDS rt-PA trial stratified by time from stroke onset and by centre, but not by stroke severity; in the 91--180 minute stratum, mean baseline NIHSS was 15.4 in the placebo arm and 13.5 in the alteplase arm (\emph{P} = 0.021)~\cite{garg2022}. An independent review committee found a clinically important and statistically significant benefit despite the imbalance~\cite{ingall2004}, and its chair said that the imbalance ``did not invalidate the results of the trial''~\cite{dyer2004}, but a 2022 re-analysis adjusting for the imbalanced covariates found that three of seven originally positive results lost significance, and argued that the imbalances more plausibly reflected allocation errors than chance~\cite{garg2022}. In the MARS trial (about 50 patients), the unadjusted survival comparison gave \emph{P} = 0.08 and the chemotherapy-alone group came out better only after adjustment for sex, histological subtype, stage and age~\cite{treasure2012} (Table 1).

\begin{table}[tbp]
\centering\small
\caption{Documented cases}
\begin{tabular}{@{}
  >{\raggedright\arraybackslash}p{(\columnwidth - 6\tabcolsep) * \real{0.2500}}
  >{\raggedright\arraybackslash}p{(\columnwidth - 6\tabcolsep) * \real{0.2500}}
  >{\raggedright\arraybackslash}p{(\columnwidth - 6\tabcolsep) * \real{0.2500}}
  >{\raggedright\arraybackslash}p{(\columnwidth - 6\tabcolsep) * \real{0.2500}}@{}}
\toprule\noalign{}
\begin{minipage}[b]{\linewidth}\raggedright
Trial
\end{minipage} & \begin{minipage}[b]{\linewidth}\raggedright
Covariate
\end{minipage} & \begin{minipage}[b]{\linewidth}\raggedright
What happened
\end{minipage} & \begin{minipage}[b]{\linewidth}\raggedright
Lesson
\end{minipage} \\
\midrule\noalign{}

MIST2~\cite{kahan2014} & Baseline effusion size (39 vs 47) & Estimate fell 38\% after adjustment; SE fell 35\% & One strong covariate can move the estimate by more than a third \\
NINDS rt-PA~\cite{garg2022,ingall2004} & Baseline NIHSS in the 91--180 min stratum & Panel upheld the result; 2022 adjustment removed significance for 3 of 7 results & Stratify on the strongest prognostic factor; imbalance can also signal allocation failure \\
MARS~\cite{treasure2012} & Sex, histology, stage, age & Chemotherapy alone better only after adjustment; unadjusted \emph{P} = 0.08 & Small trials are fragile to several modest imbalances \\
\bottomrule\noalign{}
\end{tabular}
\end{table}

Small trials report larger effects than large ones (ratio of odds ratios 0.52 for fewer than 50 versus 1000 or more patients)~\cite{dechartres2013}, as do trials with inadequate or unclear allocation concealment~\cite{schulz1995}. These findings reflect publication bias and flawed allocation; chance imbalance has a random direction and should not exaggerate effects on average unless combined with selective publication. We found no neonatal trial with a documented, consequential chance imbalance.

\hypertarget{baseline-significance-tests-answer-a-question-nobody-needs-answered}{%
\subsection{Baseline significance tests answer a question nobody needs answered}\label{baseline-significance-tests-answer-a-question-nobody-needs-answered}}

CONSORT 2025 is categorical: ``Significance testing of baseline differences is not recommended and should not be reported''~\cite{consort2025ee}. Senn called the practice ``philosophically unsound, of no practical value and potentially misleading''~\cite{senn1994}, and EMA states that such testing ``has no role'' where randomisation and blinding were ``fully satisfactory''~\cite{ema2015}. Under proper randomization the null hypothesis of a balance test is true by construction, and a \emph{P} value mixes the size of an imbalance with the sample size and contains nothing about β, so bias can be large when \emph{P} is large and negligible when \emph{P} is small~\cite{altman1985,roberts1999,deboer2015}. The arguments were set out at length in the early 1990s~\cite{altmandore1990,permutt1990,begg1990,senn1994} and restated for randomized experiments in general by Mutz and colleagues~\cite{mutz2019}. The practice persists: 38\% of trials published in four leading journals in 2007 still tested baseline characteristics~\cite{austin2010}.

Baseline \emph{P} values have one legitimate use: as a check on the randomization procedure itself, read across many allocations. Senn writes that ``tests of baseline balance are nothing if not tests of the randomisation procedure itself''~\cite{senn2017} (see also~\cite{senn1994,berger1999}), and Carlisle used the distribution of baseline \emph{P} values in 5087 published trials to flag baseline data inconsistent with random sampling~\cite{carlisle2017}. Simulating many allocations under a scheme is exactly such a procedure-level reading, and it is the only reading on which the scores discussed below rely.

\hypertarget{each-balancing-design-buys-balance-with-predictability-and-analysis-obligations}{%
\subsection{Each balancing design buys balance with predictability and analysis obligations}\label{each-balancing-design-buys-balance-with-predictability-and-analysis-obligations}}

Every restriction on randomization trades one risk for another. Accidental bias arises from an unmeasured covariate that happens to correlate with the assignment sequence, and complete randomization minimizes its worst case; selection bias arises from staff predicting the next assignment, and highly predictable balancing rules maximize it~\cite{efron1971,lachin1988a,markaryan2010}. Stronger balancing reduces covariate imbalance but raises predictability, and balance does not guarantee efficiency in nonlinear or heteroscedastic models~\cite{rosenberger2008} (Table 2).

\FloatBarrier
\begin{longtable}[]{@{}
  >{\raggedright\arraybackslash}p{(\columnwidth - 6\tabcolsep) * \real{0.2500}}
  >{\raggedright\arraybackslash}p{(\columnwidth - 6\tabcolsep) * \real{0.2500}}
  >{\raggedright\arraybackslash}p{(\columnwidth - 6\tabcolsep) * \real{0.2500}}
  >{\raggedright\arraybackslash}p{(\columnwidth - 6\tabcolsep) * \real{0.2500}}@{}}
\caption{Randomization schemes compared}\tabularnewline
\toprule\noalign{}
\begin{minipage}[b]{\linewidth}\raggedright
Scheme
\end{minipage} & \begin{minipage}[b]{\linewidth}\raggedright
What it balances
\end{minipage} & \begin{minipage}[b]{\linewidth}\raggedright
Main strength
\end{minipage} & \begin{minipage}[b]{\linewidth}\raggedright
Main costs or failure mode
\end{minipage} \\
\midrule\noalign{}
\endfirsthead
\toprule\noalign{}
\begin{minipage}[b]{\linewidth}\raggedright
Scheme
\end{minipage} & \begin{minipage}[b]{\linewidth}\raggedright
What it balances
\end{minipage} & \begin{minipage}[b]{\linewidth}\raggedright
Main strength
\end{minipage} & \begin{minipage}[b]{\linewidth}\raggedright
Main costs or failure mode
\end{minipage} \\
\midrule\noalign{}
\endhead
\bottomrule\noalign{}
\endlastfoot
Complete randomization & Nothing by design & No predictability; minimal worst-case accidental bias~\cite{markaryan2010,lachin1988b} & Chance imbalance; arm-size imbalance negligible only above about 200 patients~\cite{lachin1988b} \\
Permuted blocks & Arm sizes & Near-equal arms at all times & Selection-bias potential in unmasked trials~\cite{lachin1988a,lachin1988b} \\
Stratified permuted blocks & Arm sizes within strata & Simple and standard~\cite{ciolino2019} & Fails when cells exceed about n/2; must adjust for strata~\cite{therneau1993,kahan2012sim} \\
Efron biased coin & Arm sizes & Tunable; excess correct-guess probability 0.125 with coin bias 2/3~\cite{markaryan2010} & No covariate control unless extended \\
Wei urn & Arm sizes; margins in the covariate version & Strong early pull, relaxing towards complete randomization~\cite{wei1978} & Most predictable early, when the pull towards balance is strongest~\cite{wei1978}; margins only~\cite{rosenberger2008} \\
Biased coin on balance \emph{P} values (Frane) & The worst covariate balance \emph{P} value & Bias graded by the size of the imbalance; continuous and categorical covariates~\cite{frane1998} & Analysis by randomization test under the same constraints~\cite{frane1998} \\
Minimization & Marginal totals of many factors & Handles 10--20 factors; good small-trial balance~\cite{therneau1993,scott2002} & Predictable if deterministic; within-stratum imbalance can grow~\cite{ema2015,hu2012} \\
Rerandomization & Joint covariate distance, or every balance \emph{P} value & Unbiased under a symmetric criterion; known variance reduction~\cite{morgan2012,zhaoding2024rep} & Requires all covariates before allocation \\
Covariate-adaptive (Hu and Hu) & Overall, marginal and within-stratum imbalance & All levels bounded in probability~\cite{hu2012} & More complex; limited use \\
\end{longtable}

Stratification has a ceiling. ICH E9 states that ``the use of more than two or three stratification factors is rarely necessary''~\cite{iche9}, and Therneau found that stratified assignment ``begins to fail if the total number of distinct combination of factor levels is greater than approximately n/2,'' whereas minimization ``can accommodate a large number of factors (10--20) without difficulty''~\cite{therneau1993}. With 2\textsuperscript{10} strata and n = 500, mean absolute overall imbalance was 17.07 for stratified permuted blocks and 0.76 for Pocock--Simon minimization~\cite{hu2012}.

Minimization balances margins, not cells~\cite{taves1974,pocock1975}: EMA notes that ``dynamic allocation does not guarantee balance within combinations of prognostic factors''~\cite{ema2015}, and in a 2 × 2-strata simulation the SD of within-stratum imbalance rose from 3.16 at n = 200 to 7.25 at n = 1000~\cite{hu2012}. EMA states that ``deterministic schemes should be avoided''~\cite{ema2015}, as ICH E9 does for ``deterministic dynamic allocation procedures''~\cite{iche9}, and Coart and colleagues treat stochastic minimization as good practice provided the algorithm and allocation history are concealed~\cite{coart2023}. Frane's covariate-adaptive biased coin measures imbalance directly with balance \emph{P} values: for each candidate arm it computes every covariate's t-test, ANOVA or frequency-table \emph{P} value as if the patient were assigned there, and assigns with probability proportional to the smallest, a rule that ``maximizes the minimum between-group p-values among all prognostic variables''~\cite{frane1998}. The analysis, an ANCOVA on all balancing covariates and blocking factors, was checked by a randomization test under the same constraints; in trials of 71 and 98 patients, 4.87\% and 5.03\% of rerandomized ANCOVA \emph{P} values fell below 0.05~\cite{frane1998}. Rerandomization has the cleanest theory: accepting an allocation only when the Mahalanobis distance between arm means is at most a reduces the variance of each covariate mean difference by 100(1 − v\textsubscript{a})\%, with v\textsubscript{a} = Pr(χ\textsuperscript{2}\textsubscript{k+2} ≤ a)/Pr(χ\textsuperscript{2}\textsubscript{k} ≤ a), and the variance of the treatment estimate by 100(1 − v\textsubscript{a})R\textsuperscript{2}\%, so that the gain is capped by prognostic strength~\cite{morgan2012}. Zhao and Ding analysed rerandomization rules that accept an allocation only when every covariate's balance-test \emph{P} value exceeds a threshold, generalizing a rule used in field experiments: such rules improve balance and the efficiency of the unadjusted and additive-regression estimators, and standard regression inference remains valid but can be conservative~\cite{zhaoding2024rep}.

Centre is the most common balancing factor~\cite{kahan2012bmj}. With many small sites, stratifying by site quickly meets the n/2 limit; treating site as one factor in minimization avoids this, and the analysis should account for centre whatever is done at design~\cite{iche9,kahan2013centre,senn1998}. Balance at interim looks is a further reason for restricted designs~\cite{kernan1999,coart2023}, but under minimization within-stratum imbalance keeps growing as accrual continues, so balance at one look need not hold at the next~\cite{hu2012}, and we found no peer-reviewed study comparing covariate imbalance at interim looks across schemes.

\hypertarget{regulators-agree-on-a-chain}{%
\subsection{Regulators agree on a chain}\label{regulators-agree-on-a-chain}}

The regulatory and reporting documents agree on three links: choose a few strongly prognostic covariates in advance, balance them by design with a random element, and include the same factors in the pre-specified analysis (Table 3)~\cite{fda2023,ema2015,iche9,consort2025ee}. EMA ties imbalance most directly to credibility: post hoc imbalances ``are of particular concern if they favour the experimental group,'' and for severe imbalance ``no adjustment may be sufficiently convincing to restore the reliability of the results''~\cite{ema2015}. CONSORT 2025 also notes that chance mismatch, ``particularly in small trials,'' ``weakens the trial's credibility''~\cite{consort2025ee}. The FDA frames covariates as a tool for efficiency and says nothing explicit about chance imbalance or baseline testing~\cite{fda2023}. ICH E9 notes that stratification ``has greater potential benefit in small trials''~\cite{iche9}. No regulator or reporting guideline endorses a quantitative imbalance metric or threshold for evaluating randomization schemes at the design stage.

\FloatBarrier
\begin{longtable}[]{@{}
  >{\raggedright\arraybackslash}p{(\columnwidth - 8\tabcolsep) * \real{0.2000}}
  >{\raggedright\arraybackslash}p{(\columnwidth - 8\tabcolsep) * \real{0.2000}}
  >{\raggedright\arraybackslash}p{(\columnwidth - 8\tabcolsep) * \real{0.2000}}
  >{\raggedright\arraybackslash}p{(\columnwidth - 8\tabcolsep) * \real{0.2000}}
  >{\raggedright\arraybackslash}p{(\columnwidth - 8\tabcolsep) * \real{0.2000}}@{}}
\caption{What the regulatory and reporting documents say}\tabularnewline
\toprule\noalign{}
\begin{minipage}[b]{\linewidth}\raggedright
Topic
\end{minipage} & \begin{minipage}[b]{\linewidth}\raggedright
FDA 2023
\end{minipage} & \begin{minipage}[b]{\linewidth}\raggedright
EMA 2015
\end{minipage} & \begin{minipage}[b]{\linewidth}\raggedright
ICH E9
\end{minipage} & \begin{minipage}[b]{\linewidth}\raggedright
CONSORT 2025
\end{minipage} \\
\midrule\noalign{}
\endfirsthead
\toprule\noalign{}
\begin{minipage}[b]{\linewidth}\raggedright
Topic
\end{minipage} & \begin{minipage}[b]{\linewidth}\raggedright
FDA 2023
\end{minipage} & \begin{minipage}[b]{\linewidth}\raggedright
EMA 2015
\end{minipage} & \begin{minipage}[b]{\linewidth}\raggedright
ICH E9
\end{minipage} & \begin{minipage}[b]{\linewidth}\raggedright
CONSORT 2025
\end{minipage} \\
\midrule\noalign{}
\endhead
\bottomrule\noalign{}
\endlastfoot
Baseline significance tests & Not addressed & ``No role'' if randomisation and blinding were ``fully satisfactory'' & No recommendation to test & ``not recommended and should not be reported'' \\
Chance imbalance found afterwards & Not addressed explicitly & Not a reason for primary adjustment; ``of particular concern if they favour the experimental group'' & Supportive analysis accounting for it & Judge by prognostic strength and magnitude \\
Choice of covariates & Those most strongly associated with outcome & Strong a priori association, justified externally; ``no more than a few'' & Pre-specified in the protocol & Rationale and pre-specification reported \\
Balancing factors in the analysis & Model ``should generally include strata variables'' & Usually included ``regardless of their prognostic value'' & ``should be accounted for later in the analysis'' & Adjustment for stratification variables often advisable \\
Dynamic allocation & Discuss complex covariate-adaptive randomization with the review division & ``Deterministic schemes should be avoided'' & Include ``an appropriate element of randomisation'' & Identify minimization and its variables \\
Non-collapsibility & Defined; choose the estimand in advance & Adjusted and unadjusted effects differ in interpretation in non-linear models & Not covered & Not covered \\
\end{longtable}

Sources:~\cite{fda2023,ema2015,iche9,consort2025ee,consort2010}.

\hypertarget{calibrating-balance-p-values-for-design-stage-scoring}{%
\subsection{\texorpdfstring{Calibrating balance \emph{P} values for design-stage scoring}{Calibrating balance P values for design-stage scoring}}\label{calibrating-balance-p-values-for-design-stage-scoring}}

We now consider a design-stage score built on the worst balance \emph{P} value over a battery of covariate tests, read across many simulated allocations, with \emph{P} ≤ 1/e ≈ 0.368 flagged as crossing the pre-specified balance threshold. The score is a property of a scheme, not of any trial. Four properties govern how it should be read.

\emph{Why 1/e.} For \emph{P} \textless{} 1/e, B(\emph{P}) = −e·\emph{P}·ln(\emph{P}) lower-bounds the Bayes factor for the null against the alternative; for \emph{P} ≥ 1/e the bound is 1~\cite{sellke2001}. Bayarri and colleagues, who propose reporting the bound in practice, state it for \emph{P} ≤ 1/e ≈ 0.37~\cite{bayarri2016}, and Held and Ott show where the cut-off comes from: if the \emph{P} value under the alternative follows a Beta(ξ, 1) distribution, the maximum-likelihood estimate of ξ is min\{−1/ln \emph{P}, 1\}, which reaches 1, the uniform distribution of the null, at \emph{P} = 1/e, so that beyond it the minimum Bayes factor is 1~\cite{heldott2018}. Read as maximum odds for the alternative, 1/B(\emph{P}), a \emph{P} value of 0.05 carries at most 2.46 to 1 and a \emph{P} value of 1/e exactly even odds (Figure 4a; Additional file 1, Table A5)~\cite{sellke2001,benjamin2019,vovk1993}. On this scale, 1/e is the point at which a \emph{P} value stops carrying any possible evidence against the null. The same point appears in Fisher's method of combining \emph{P} values. Each \emph{P} value adds −2 ln \emph{P} to a chi-square statistic and two degrees of freedom to its reference distribution, and the addition equals its null expectation of 2 exactly at \emph{P} = 1/e (\emph{derived}; D7). Fisher noted that a \emph{P} value above one half, because it introduces ``two additional degrees of freedom'', ``will tend to lower'' the significance of the combined test~\cite{fisher1948}. Elston showed that a \emph{P} value of e\textsuperscript{−1} ≈ 0.37, replicated many times, gives a combined \emph{P} value near 0.5, and concluded that, if one accepts the weight-free ranking of \emph{P} value vectors from which Fisher's method follows, ``the value p = 0.37 is a critical one''~\cite{elston1991}. Two caveats apply. The calibration assumes a precise null and a real alternative, whereas under randomization the balance null is true by design; it is used here as a scale of how unremarkable an allocation looks, not as a Bayes factor. And our review found no independent published source that ties 1/e to baseline balance; the nearest proposal, Ciolino and colleagues' one-sided level of about 24\% (about 0.48 two-sided), came from simulated loss of power rather than from a calibration~\cite{ciolino2015}. The use of 1/e for balance scoring is a design choice; Property 4 shows that, for a single balance \emph{P} value, it is also a calibrated one.

\emph{Property 1: the score measures improvement over complete randomization.} For a test whose null \emph{P} values are continuous and exactly calibrated, every balance \emph{P} value is Uniform(0, 1) under complete randomization, whatever n and whatever the covariate's prognostic strength; Senn proposed describing imbalance by its \emph{P} value for exactly this reason, since ``under randomization this has a uniform distribution between 0 and 1, whatever the size of the sample''~\cite{senn1989}. The worst of k independent \emph{P} values therefore satisfies

\[\Pr(\text{worst } P \le 1/e) = 1 - (1 - 1/e)^k, \qquad (7)\]

which is 0.37 for one covariate and 0.99 for ten (Figure 4b; Table A6; \emph{derived}; D8). The threshold is deliberately strict: only 1\% of completely randomized allocations keep all ten independent \emph{P} values at or above 1/e. A score is therefore interpretable only against the complete-randomization benchmark for the same covariate set. For jointly normal balance statistics and two-sided tests, correlation can only lower the benchmark (Šidák's inequality; D8), and exact tests on categorical covariates give \emph{P} values that are not exactly uniform (Fisher's exact test is conservative; its mid-\emph{P} version is closer to uniform), so the benchmark should be simulated rather than assumed (Additional file 1, Figure A5).

\begin{figure}
\centering
\includegraphics[width=1\textwidth,height=\textheight]{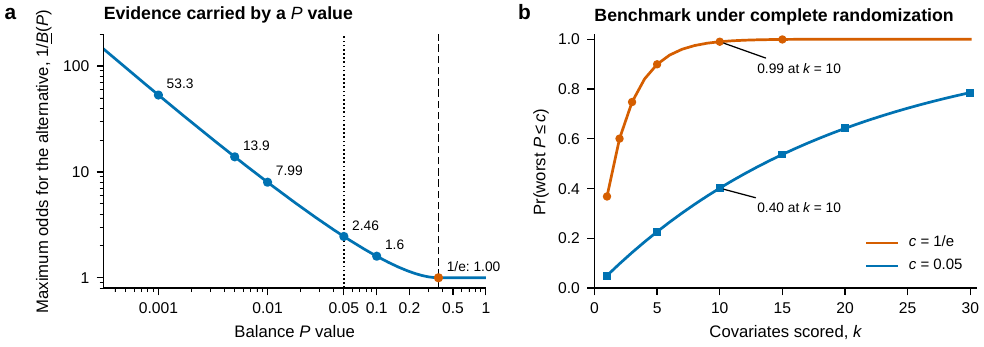}
\caption{\textbf{Why 1/e, and the benchmark it sets.} (\textbf{a}) Maximum odds for the alternative carried by a \emph{P} value, 1/B(\emph{P}), from the Sellke--Bayarri--Berger bound, on logarithmic scales; points give the values in Additional file 1, Table A5. The dashed line marks \emph{P} = 1/e, where the bound reaches even odds and stays there; the dotted line marks \emph{P} = 0.05. Used here as a scale of how unremarkable an allocation looks, not as a Bayes factor about a real alternative. (\textbf{b}) Probability that the worst of k independent, continuous and exactly calibrated balance \emph{P} values falls at or below a threshold c under complete randomization, Pr(worst \emph{P} ≤ c) = 1 − (1 − c)\textsuperscript{k} (equation 7), for c = 1/e (circles; Additional file 1, Table A6) and c = 0.05 (squares; Tables A3 and A6). A scheme's score is interpretable only against this benchmark for the same covariates; for jointly normal two-sided statistics, correlated covariates lower it (Additional file 1, Figure A5). Additional file 1, D7 and D8.}
\end{figure}

\emph{Property 2: a balance P value maps to bias only through prognostic strength.} From (6), with \textbar Z\textsubscript{bal}\textbar{} = Φ\textsuperscript{−1}(1 − \emph{P}/2), the conditional bias of the unadjusted estimate in standard errors is ρΦ\textsuperscript{−1}(1 − \emph{P}/2) (Figure 5; Table A7; \emph{derived}; D9). At 1/e, \textbar Z\textsubscript{bal}\textbar{} = 0.90, so the bias ranges from 0.27 SE at ρ = 0.3 to 0.81 SE at ρ = 0.9. A balance \emph{P} of 0.05 is harmless for a covariate with ρ near zero and shifts the estimate by 1.76 SE when ρ = 0.9. This is the formal version of CONSORT's instruction to weigh ``prognostic strength'' and ``magnitude'' together~\cite{consort2025ee}: a score says nothing about bias until each covariate carries a prognostic weight taken from external data, as EMA requires for covariate selection~\cite{ema2015}.

\begin{figure}
\centering
\includegraphics[width=0.85\textwidth,height=\textheight]{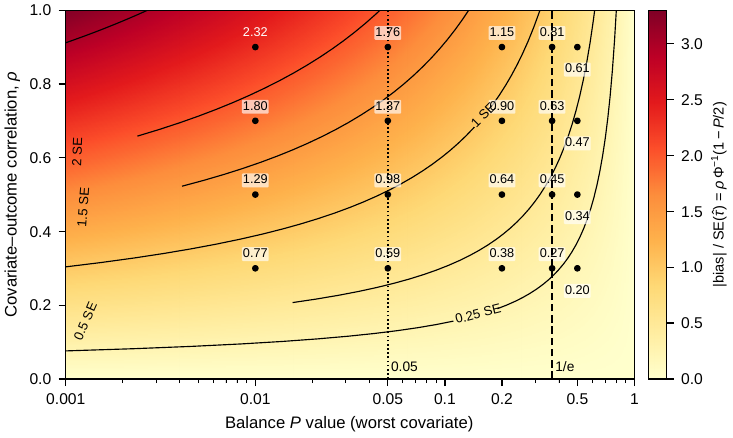}
\caption{\textbf{Conditional bias implied by a balance \emph{P} value depends on prognostic strength.} Colour gives the absolute conditional bias of the unadjusted estimate in standard errors, ρΦ\textsuperscript{−1}(1 − \emph{P}/2) (equation 6), as a function of the balance \emph{P} value and the covariate--outcome correlation ρ; black contours at 0.25 to 3 standard errors. Printed values are the entries of Additional file 1, Table A7. The dashed line marks \emph{P} = 1/e and the dotted line \emph{P} = 0.05. The same balance \emph{P} value implies a negligible bias for a weak covariate and a bias of more than 1.5 standard errors for a strong one. Additional file 1, D9.}
\end{figure}

\emph{Property 3: for one covariate, the 1/e threshold controls two-sided but not fully one-sided conditional error.} Under H\textsubscript{0} and model (1) with normal data, (Z\textsubscript{bal}, T\textsubscript{u}) is bivariate standard normal with correlation ρ, so

\[T_u \mid Z_{\text{bal}} \sim N\!\left(\rho Z_{\text{bal}},\; 1 - \rho^2\right) \qquad (8)\]

(\emph{derived}; D10). This is the model Senn used in 1989 to study the effect of imbalance on the size of the test of efficacy~\cite{senn1989}. Working with a one-sided test, he derived the conditional size in the direction the imbalance favours, the imbalance at which it first exceeds nominal and, for a given imbalance, the correlation that makes it largest, and concluded that tests of baseline homogeneity cannot control size. Permutt wrote down the two-sided conditional rejection probability of the unadjusted test in 1990 and showed that it rises with the size of the imbalance and averages to the nominal level, so that adjusting only after a significant balance test gives a procedure whose size is below nominal and whose power falls well short of always adjusting when the correlation exceeds about 0.4~\cite{permutt1990}. In 1994 Senn gave the two-sided conditional size for arms of any size, and showed by example that a covariate with ρ = 0.73 and a balance \emph{P} value of 0.15, which passes a 5\% balance test, nearly doubles the size of a 5\% test of efficacy, to 0.09~\cite{senn1994}. Senn also gave analytic one-sided conditional sizes for crude outcomes, change from baseline and ANCOVA~\cite{senn1993}, Overall and Magee tabulated the directional error of several analyses by simulation in bands of the baseline \emph{P} value~\cite{overall1992}, and Zhao and Ding have since studied preliminary balance testing in a randomization-based framework~\cite{zhaoding2024pre}. The boundary analysis that follows is, to our knowledge, new. Expanding the two-sided conditional size for small ρ gives α + qφ(q)ρ\textsuperscript{2}(Z\textsubscript{bal}\textsuperscript{2} − 1) + O(ρ\textsuperscript{4}), with q = 1.96. To second order, the size therefore exceeds nominal exactly when \textbar Z\textsubscript{bal}\textbar{} \textgreater{} 1, that is, when the balance \emph{P} value falls below 2Φ(−1) = 0.3173\ldots. At a known ρ the crossing lies at smaller \emph{P}: 0.301 at ρ = 0.5, 0.262 at ρ = 0.7 and 0.167 at ρ = 0.9 (D10; Figure 6a). Across all ρ the condition is exact: any balance \emph{P} value at or above 2Φ(−1) keeps the two-sided conditional type I error of the unadjusted test at or below 5\% for every ρ, and no smaller cut-off does, so a rounded value such as 0.317 is not sufficient (\emph{derived}; D10, which gives the proof). 1/e lies inside that region. This is an independent, frequentist reason the threshold is sensible, although it was not the reason for choosing it.

The one-sided picture, Senn's case, is less reassuring. False positives in the direction favoured by the imbalance exceed 2.5\% whenever ρZ\textsubscript{bal} \textgreater{} q\{1 − √(1 − ρ\textsuperscript{2})\}, Senn's boundary~\cite{senn1989} (Figure 6b; Table A8). At 1/e the directional error is at most 4.1\%, reached at ρ = \textbar Z\textsubscript{bal}\textbar/q ≈ 0.46, the worst-case correlation of Senn's analysis, against 13--33\% at a balance \emph{P} of 0.05 for ρ between 0.5 and 0.9. Because EMA singles out imbalance that favours the experimental arm~\cite{ema2015}, this directional view is the more relevant one for regulatory credibility. At very high ρ and modest imbalance the conditional error falls below nominal, because the unadjusted test's variance includes covariate variation that conditioning has removed; Senn noted the same for two-sided tests~\cite{senn1989}. The unadjusted test is miscalibrated in both directions.

\begin{figure}
\centering
\includegraphics[width=1\textwidth,height=\textheight]{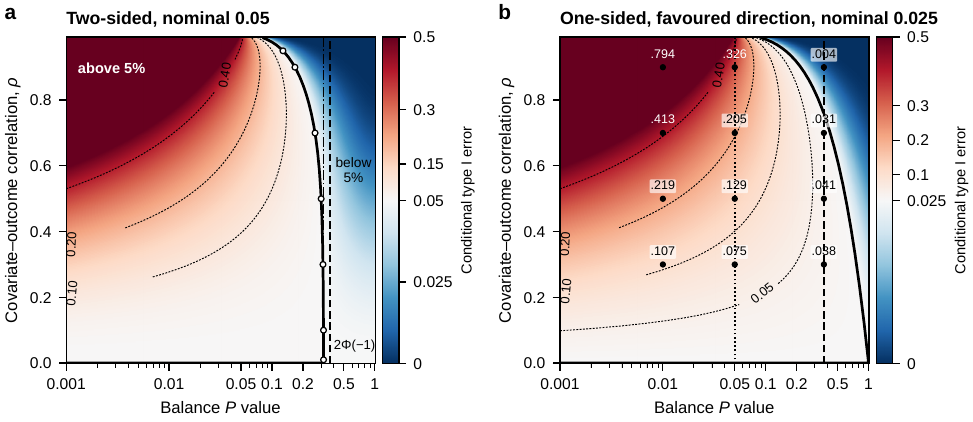}
\caption{\textbf{Conditional type I error of the unadjusted test given the balance \emph{P} value.} Under the null hypothesis and model (1), from equation (8). (\textbf{a}) Two-sided error at nominal 0.05: red above nominal, blue below. The thick black contour is the boundary at which the error equals 0.05, with open circles at the exact values tabulated in Additional file 1, D10; the dash-dot line marks \emph{P} = 2Φ(−1) and the dashed line \emph{P} = 1/e. Every balance \emph{P} value of at least 2Φ(−1) lies at or below nominal for all ρ (equality at ρ = 0). (\textbf{b}) One-sided error in the direction favoured by the imbalance at nominal 0.025, with the values of Additional file 1, Table A8 printed for \emph{P} = 0.01, 0.05 and 1/e; the thick contour marks 0.025. Thin dotted contours give further error levels. Additional file 1, D10.}
\end{figure}

\emph{A precision reading of the threshold.} If complete-randomization allocations were retained only when a single normal covariate's balance \emph{P} value is at least 1/e (rerandomization with that criterion), the variance of its standardized difference would fall to 24.2\% of its value under complete randomization, which equals Morgan and Rubin's v\textsubscript{a} for k = 1~\cite{morgan2012}; the variance of the unadjusted treatment estimate would fall by 75.8\% × R\textsuperscript{2} (\emph{derived}; D11; Additional file 1, Figure A9). Sequential schemes do not truncate the distribution this cleanly, but each scheme's simulated distribution can be read the same way.

\emph{Property 4: for a single balance P value, a local Beta shift gives the cut-off 1/e, and a local normal contraction gives 2Φ(−1).} Because a score is used to compare schemes, a natural criterion for a cut-off c is how far it separates a scheme's probability of a flag from that under complete randomization, D(c) = c − F(c), where F is the scheme's distribution function for the balance \emph{P} value. Two local models of improvement give closed-form answers. If the scheme shifts \emph{P} values upward within the Beta(1 + δ, 1) family, so that F(c) = c\textsuperscript{1+δ}, then D(c) = δ(−c ln c) + O(δ\textsuperscript{2}), whose first-order term is maximized uniquely at c = 1/e; at a finite improvement the maximizer is (1 + δ)\textsuperscript{−1/δ}, which exceeds 1/e and reaches 1/2 at δ = 1 (\emph{derived}; D13). This is the family of Held and Ott's derivation~\cite{heldott2018}, taken in the opposite direction: the Sellke--Bayarri--Berger bound, −e·\emph{P}·ln(\emph{P}), is built from the same function, but its alternatives favour small \emph{P} values and a better-balancing scheme favours large ones, so the two calibrations share a function and its maximum, not an interpretation. If instead the scheme contracts the balance statistic, so that Z\textsubscript{bal} \ensuremath{\sim} N(0, (1 − ε)\textsuperscript{2}), the local gain is 2εaφ(a) with a = Φ\textsuperscript{−1}(1 − c/2), maximized at a = 1, that is at c = 2Φ(−1), the two-sided boundary of Property 3 (\emph{derived}; D13). Each of the two cut-offs retains about 99\% of the other model's best local separation, and every cut-off from 0.30 to 0.40 retains at least 97\% under both. For a single balance \emph{P} value, then, the Bayes-factor calibration, the error boundary and the separation criterion converge on the band from 2Φ(−1) to 1/e. For finite improvements the largest gap in this Lehmann family was evaluated by Wellek~\cite{wellek2021}; its local limit, 1/e, is the cut-off used here, and what it adds is a reason for the threshold that does not rest on the Bayes-factor scale. It does not carry over to the worst of k covariates. The local gain then acquires the factor k(1 − c)\textsuperscript{k−1}, and for k independent, equally improved covariates the best cut-off falls to about 0.24 (Beta) and 0.21 (normal) for k = 2, and to 0.12 and 0.11 for k = 5, where 1/e retains 38\% and 34\% of the attainable separation (D13). For a multi-covariate score, 1/e is therefore a fixed, calibrated per-covariate reference, read against the complete-randomization benchmark of Property 1, not an optimal discriminator.

The error results (Properties 2 and 3) concern one normally distributed covariate, a large sample and an unadjusted Gaussian outcome test; the separation result (Property 4) concerns the design distribution of a continuous balance \emph{P} value. With several prognostic covariates the total shift is a weighted sum of their z-statistics, and per-covariate cut-offs bound each imbalance but can allow a standardized prognostic shift greater than one. Two independent covariates with balance z-statistics of 0.88 each have \emph{P} = 0.379, above 1/e; if both are prognostic in the same direction with equal standardized weights and their composite has correlation 0.5 with the outcome, the two-sided conditional size of the unadjusted test is 6.3\%, against 2.4\% when the same imbalances point in opposite directions, so no rule that sees only the two \emph{P} values can tell these allocations apart (\emph{derived}; D13). A guarantee that holds for every prognostic direction and strength, in this joint normal model with known covariance, requires the Mahalanobis distance of the balance statistics to be at most 1, in effect a rerandomization criterion~\cite{morgan2012} (D13). Under a pre-specified ANCOVA on the same covariates, conditional bias is removed and the guarantees are not needed for validity, provided the adjustment model is right: Maxwell showed that ANCOVA's control of conditional size depends on the accuracy of the model, and that stratified assignment lessens that dependence~\cite{maxwell1993}, which is one reason to balance strong prognostic factors by design even when the analysis will adjust for them. Imbalance then costs only precision: for one covariate and equal arms, the variance of the ANCOVA estimate is inflated by exactly 1 + t\textsubscript{bal}\textsuperscript{2}/(2n − 2), where t\textsubscript{bal} is the pooled-variance balance t statistic, which is the variance inflation factor of Senn, König and Posch~\cite{senn2026}. Under complete randomization its expectation is 1 + 1/(2n − 4); at a balance \emph{P} value of 1/e the inflation is 2.2\% with 20 patients per arm and 0.4\% with 100 (\emph{derived}; D3.5).

\hypertarget{schemes-compared-on-the-score}{%
\subsection{Schemes compared on the score}\label{schemes-compared-on-the-score}}

To see what the score distinguishes in practice, we simulated 20 000 allocations of the same patients under six schemes, with targets of 25, 50 and 100 patients per arm, with five covariates scored (three continuous and two binary, with prevalences 0.30 and 0.50 and pairwise latent correlation 0.2), an outcome driven by a prognostic index of the five covariates with correlation ρ of 0.3, 0.5 or 0.7, and no treatment effect (Additional files 1, Section N, and 4). Predictability was the probability of guessing the next assignment correctly by choosing the arm that was behind overall, behind within the patient's stratum for stratified blocks, or preferred by the rule for minimization. Minimization split the first continuous covariate at its pre-specified population median, so that every patient's factor levels were known at enrolment. Under complete randomization the simulated benchmark agreed with equation (7) to within Monte Carlo error for one to ten independent covariates (0.8990 against 0.8991 for five), stayed between 0.897 and 0.902 from 10 to 500 patients per arm, and fell to 0.890 with correlation 0.2 and to 0.836 with correlation 0.5; with two of the five covariates binary it was 0.892.

The schemes separated in three ways (Figure 7, 50 patients per arm). Permuted blocks and Efron's biased coin had scores similar to that of complete randomization (0.887 and 0.895 against 0.889), because they balance arm sizes rather than covariates, yet they raised the probability of a correct guess from 0.56 to 0.71 and 0.62. Stratified blocks on the two binary covariates and minimization on three factors lowered the score to 0.72 and 0.66, at correct-guess probabilities of 0.70 and 0.74. Rerandomization, shown only as a reference because it needs every patient's covariates in advance, reached 0.03; with the same count-based guesser its predictability equalled that of complete randomization, which is not the best prediction open to someone who knows the covariates and the acceptance rule. The score acted on the unadjusted test through ρ, as Property 2 predicts: across schemes, the two-sided error of the unadjusted test ranged from 0.0515 to 0.0197 (0.0112 with rerandomization) at ρ = 0.7, but only from 0.0512 to 0.0421 (0.0411) at ρ = 0.3. Under complete randomization at ρ = 0.7, 1014 of the 1030 false positives (98\%) lay in the direction favoured by the prognostic imbalance, a rate of 0.0507 against 0.025 if the imbalance were unrelated to the outcome; this is Senn's one-sided problem seen across trials. Pre-specified ANCOVA on the five covariates kept its type I error between 0.046 and 0.052 under every scheme and sample size; in this design its rates do not depend on ρ. What balancing bought a correctly adjusted analysis was precision, and little of it: the ANCOVA standard error was 1.9\% to 2.3\% smaller than under complete randomization with stratified blocks or minimization at 25 patients per arm, 1.0\% to 1.2\% at 50 and 0.5\% to 0.6\% at 100, and 4.6\% smaller with rerandomization at 25.

\begin{figure}
\centering
\includegraphics[width=1\textwidth,height=\textheight]{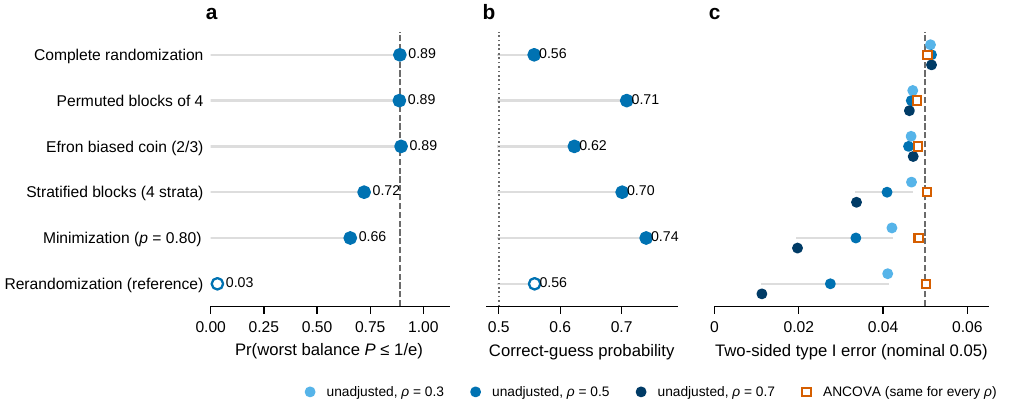}
\caption{\textbf{Six randomization schemes compared on the design-stage score, predictability and type I error.} Simulated allocations (20 000 per scheme) with a target of 50 patients per arm, five covariates scored (three continuous, two binary; latent correlation 0.2) and no treatment effect. (\textbf{a}) Score: probability that the worst of the five balance \emph{P} values is at most 1/e; the dashed line marks complete randomization. (\textbf{b}) Probability of guessing the next assignment correctly by choosing the arm that is behind overall, behind within the patient's stratum (stratified blocks) or preferred by the rule (minimization); the dotted line marks 0.5. (\textbf{c}) Two-sided type I error of the unadjusted t test at covariate--outcome correlation ρ = 0.3, 0.5 and 0.7 (circles), and of pre-specified ANCOVA on all five covariates (squares; the same for every ρ in this design); the dashed line marks nominal 0.05. Minimization splits the first continuous covariate at its pre-specified population median. Rerandomization (open circles) is a reference only: it needs every patient's covariates before allocation, and its predictability is measured with the same count-based guesser as complete randomization. Additional file 1, Section N; Additional file 4.}
\end{figure}

\hypertarget{what-a-balance-score-cannot-see}{%
\subsection{What a balance score cannot see}\label{what-a-balance-score-cannot-see}}

A balance score has blind spots that should be reported alongside it (Table 4).

\begin{table}[tbp]
\centering\small
\caption{Blind spots of a design-stage balance score, and proposed items to report with it}
\begin{tabular}{@{}
  >{\raggedright\arraybackslash}p{(\columnwidth - 4\tabcolsep) * \real{0.3333}}
  >{\raggedright\arraybackslash}p{(\columnwidth - 4\tabcolsep) * \real{0.3333}}
  >{\raggedright\arraybackslash}p{(\columnwidth - 4\tabcolsep) * \real{0.3333}}@{}}
\toprule\noalign{}
\begin{minipage}[b]{\linewidth}\raggedright
Blind spot
\end{minipage} & \begin{minipage}[b]{\linewidth}\raggedright
Why the score misses it
\end{minipage} & \begin{minipage}[b]{\linewidth}\raggedright
Report alongside the score
\end{minipage} \\
\midrule\noalign{}

Prognostic strength & \emph{P} values contain no information about β & Prognostic weights from external data; bias in SE units (Figure 5) \\
Non-collapsibility & Present under perfect balance & The planned estimand and analysis model \\
Unmeasured covariates & Only measured covariates are scored & Accidental-bias exposure of the scheme~\cite{markaryan2010} \\
Predictability and selection bias & Balance and predictability move together~\cite{rosenberger2008} & A predictability metric, such as expected correct guesses \\
Analysis plan & Balance pays off only if balancing factors are adjusted for~\cite{kahan2012sim,morris2022} & A statement that the analysis includes all balancing factors and site \\
Within-stratum imbalance & Marginal balance can hide cell imbalance~\cite{ema2015,hu2012} & Marginal and within-stratum (and within-site) scores, separately \\
Interim looks & Within-stratum imbalance can grow after an early look~\cite{hu2012} & Scores at each planned information fraction \\
Allocation failure & Simulation assumes the scheme runs as designed~\cite{garg2022} & Concealment and audit procedures \\
\bottomrule\noalign{}
\end{tabular}
\end{table}

\hypertarget{discussion}{%
\section{Discussion}\label{discussion}}

\hypertarget{principal-findings}{%
\subsection{Principal findings}\label{principal-findings}}

Randomization protects the average, not the trial. Given the allocation actually drawn, the unadjusted estimate is biased by β × Δ\textsubscript{X}, and that bias, measured in standard errors, equals ρ × Z\textsubscript{bal} at every sample size. Its consequences depend on prognostic strength rather than on the balance \emph{P} value, and the analysis decides its sign: final-value analysis carries it, change scores can reverse it, and a correctly specified, pre-specified ANCOVA removes it while gaining precision. Non-collapsibility is a separate, estimand-level effect that no randomization scheme can address. Regulators converge on few, pre-specified, prognostic covariates balanced by design and adjusted in analysis, and on the abandonment of baseline significance tests. In simulation, schemes that balance arm sizes did not improve the score at all; stratification and minimization improved it at the price of predictability; and no scheme changed the type I error of a pre-specified adjusted analysis, to which balancing added a few per cent of precision at most.

\hypertarget{what-is-new}{%
\subsection{What is new}\label{what-is-new}}

The contribution beyond synthesis is a set of results that connect balance \emph{P} values, read at the level of the scheme, to the bias and error rates of the eventual analysis. Its components have precedents. Permutt and Senn gave the two-sided conditional size of the unadjusted test~\cite{permutt1990,senn1994}, and Senn the one-sided conditional size, its boundary, its independence from sample size and one-sided conditional sizes for change scores and ANCOVA~\cite{senn1989,senn1993}; Overall and Magee simulated the directional error of several analyses by band of baseline \emph{P} value~\cite{overall1992}; Ciolino and colleagues used the balance t statistic as a measure of imbalance and tabulated the values at which unadjusted power and size deteriorate~\cite{ciolino2015}; Morgan and Rubin derived the precision gained from a design threshold~\cite{morgan2012}; Frane built an allocation rule on the worst balance \emph{P} value~\cite{frane1998}; and Zhao and Ding gave randomization-based theory for rerandomization on balance-test \emph{P} values and for preliminary balance testing~\cite{zhaoding2024rep,zhaoding2024pre}. What we add is narrower. First, for a single covariate, a balance \emph{P} value of at least 2Φ(−1) = 0.3173\ldots{} keeps the two-sided conditional error of the unadjusted test at or below nominal for every ρ, and the cut-off is sharp, so one threshold suffices without knowing prognostic strength (Figure 6; proof in Additional file 1, D10). Second, the worst-of-k benchmark (equation 7, Figure 4b), checked by simulation, is the reference against which any scheme-level score must be read. Third, we compare common schemes by simulation on that score alongside predictability and the error rates of the unadjusted and adjusted analyses (Figure 7). Fourth, for a single balance \emph{P} value the cut-off is calibrated rather than arbitrary: 1/e maximizes the local separation between an improved scheme and complete randomization under a Beta shift, the corresponding optimum under normal contraction is the boundary 2Φ(−1), each retains about 99\% of the other's separation, and the Bayes-factor, error-control and separation criteria converge on the band between them (Property 4; D13). For worst-of-k scores over independent, equally improved covariates the locally best cut-off is lower, so there 1/e serves as a fixed reference read against the benchmark. The mapping bias/SE = ρZ\textsubscript{bal} (Figure 5) puts numbers on the relevance objection to baseline testing, and the truncation result links the threshold to Morgan and Rubin's variance reduction~\cite{morgan2012}. The choice of 1/e is calibrated, not validated: our review identified no independent validation of this threshold for detecting clinically consequential joint imbalance.

\hypertarget{relation-to-other-work}{%
\subsection{Relation to other work}\label{relation-to-other-work}}

Lachin treats imbalance as a computable property of a procedure and equates it with the potential for accidental bias~\cite{lachin1988a}. Morgan and Rubin set a design threshold that is in effect a balance-test \emph{P} value cut-off and derive its effect on precision~\cite{morgan2012}. Senn concedes baseline tests a role only as tests of the randomization procedure itself~\cite{senn1994,senn2017}, which is the role they play when a scheme's allocations are simulated. EMA links imbalance to regulatory credibility~\cite{ema2015}, and ICH E9, Kernan and colleagues, CONSORT 2025 and Coart and colleagues identify small trials and interim looks as the settings in which design balance matters most~\cite{iche9,kernan1999,consort2025ee,coart2023}. Frane came closest to the present metric: his allocation rule maximizes the worst balance \emph{P} value across covariates for each new patient~\cite{frane1998}, and Zhao and Ding's marginal rule applies the same criterion to rerandomization, accepting an allocation only when every balance \emph{P} value clears a threshold~\cite{zhaoding2024rep}. Ciolino and colleagues proposed judging a realized allocation by its balance t statistic against thresholds tied to power and size~\cite{ciolino2015}. What the literature does not supply is the use of the worst balance \emph{P} value, read over many simulated allocations, to score and compare sequential schemes against complete randomization, or the 1/e threshold; both should be presented as the designer's construction.

\hypertarget{limitations}{%
\subsection{Limitations}\label{limitations}}

This is a narrative, not a systematic, review, and relevant studies may have been missed, although every cited source was read in full text and every quotation and specific figure checked against it. The analytic results assume a linear model, one normally distributed covariate, equal arms and known variances. Replacing normal quantiles with t quantiles widens the tails slightly in very small trials, as Senn anticipated~\cite{senn1989}, but with estimated variances and t-tests the simulated conditional errors differed from the normal-theory values by at most 0.6 percentage points at 10 patients per arm and 0.2 points at 25 per arm (Additional file 1, Section M). Categorical covariates and exact tests make balance \emph{P} values discrete and non-uniform. With several correlated prognostic covariates the shifts combine, and per-covariate thresholds do not keep the combined shift within the single-covariate guarantee. The Sellke--Bayarri--Berger calibration is used as a scale rather than a Bayes factor about a real alternative; even as a bound it holds for local alternatives on a scalar parameter in large samples, but not necessarily in small samples or for parameters of higher dimension~\cite{heldott2018}. Property 4 rests on two local models of how a scheme improves balance and on one separation criterion; finite improvements, other models or criteria, and discrete tests can select other cut-offs. The scheme comparison used one covariate set, one set of prognostic weights and one setting of each scheme, and its numbers apply to that setting; rerandomization is shown only as a reference, because it needs every patient's covariates before anyone is allocated. Finally, no empirical study has linked design-stage balance scores to the credibility or accuracy of completed trials; the simulation describes schemes, not trials, and validation in a cohort of trials sampled without regard to outcome is needed.

\hypertarget{implications}{%
\subsection{Implications}\label{implications}}

For trial designers, the practical reading of a scheme score runs as follows: benchmark against complete randomization on the same covariates and sample size; judge the improvement at the final sample size and at planned interim looks; weight covariates by prognostic strength; check stratified schemes against the n/2 rule and minimization schemes for within-stratum and within-site balance; report predictability next to balance; and state that the analysis will adjust for every balancing factor, including site. Without the last step a balancing scheme yields conservative tests and lost power rather than better inference~\cite{fda2023,kahan2012sim}. In neonatal trials, where gestational age and late-onset sepsis are strongly prognostic, these considerations are especially pressing, and simulation studies of imbalance in those factors are lacking.

\hypertarget{conclusions}{%
\section{Conclusions}\label{conclusions}}

Chance imbalance biases the trial that is run, by an amount set by prognostic strength, and does not become harmless as trials grow. Correctly specified, pre-specified adjustment removes it; design can narrow its spread. A design-stage balance score estimates how often an allocation scheme crosses a pre-specified balance threshold on the covariates that matter, relative to complete randomization, and at what cost in predictability. For one normally distributed prognostic covariate, 1/e lies inside the region of uniform two-sided conditional error control and is the best, or nearly the best, cut-off for separating schemes under two local models. With several covariates, the effect of imbalance depends on their joint distribution and on the direction of their association with the outcome. The threshold is calibrated, not validated, and should be reported as such.

\hypertarget{abbreviations}{%
\section{Abbreviations}\label{abbreviations}}

ANCOVA: analysis of covariance; ANOVA: analysis of variance; CP: conditional power; EMA: European Medicines Agency; FDA: US Food and Drug Administration; HR: hazard ratio; ICC: intraclass correlation coefficient; ICH: International Council for Harmonisation; NICHD: Eunice Kennedy Shriver National Institute of Child Health and Human Development; NIHSS: National Institutes of Health Stroke Scale; OR: odds ratio; SD: standard deviation; SE: standard error.

\hypertarget{declarations}{%
\section{Declarations}\label{declarations}}

\textbf{Ethics approval and consent to participate.} Not applicable. This is a methodological review with analytic derivations and simulation; it used published literature and simulated data only and involved no human participants.

\textbf{Consent for publication.} Not applicable. No individual-person data are reported.

\textbf{Clinical trial number.} Not applicable.

\textbf{Availability of data and materials.} No new data were generated. All code, seeds and the expected output are deposited; the reproducibility package is openly archived on Zenodo; the concept DOI is \href{https://doi.org/10.5281/zenodo.24210820}{10.5281/zenodo.24210820}, which always resolves to the latest version, and this paper corresponds to v1.3, DOI \href{https://doi.org/10.5281/zenodo.24520628}{10.5281/zenodo.24520628}~\cite{dwyer2026code}. Every derived number and table, including Tables A3--A8, the D10 boundary table, the D13 separation calculations and the Monte Carlo and small-trial checks of Additional file 1 (Section M), is recomputed by imbalance\_calculations.py (Additional file 2), and Figures 1--6 and Additional file 1 Figures A1--A12 are drawn by make\_figures.py (Additional file 3) from the formulas of Additional file 1 and from fixed-seed simulations. The scheme comparison and Figure 7 are produced by scheme\_comparison.py and make\_figure7.py (Additional files 4 and 5), which are in the archive from version 1.2. The archive reproduces the study's calculations and simulations in the recorded software environment; the last decimal of a Monte Carlo value can vary with software versions. Values drawn from prior studies are attributed to their sources. A preprint is also available on Zenodo, DOI \href{https://doi.org/10.5281/zenodo.24210825}{10.5281/zenodo.24210825}.

\textbf{Competing interests.} WJD develops and hosts the open-source software and associated web domains (the trialdesign.com applications) that implement the design-stage balance scoring described. WJD was the sole inventor on US patent application 13/768,598 (publication US 2013/0211805 A1) on design-stage balance scoring~\cite{dwyer2013}; the application was abandoned in 2019. The authors use 1/e as a pre-specified per-covariate reference. Its calibration for the single-covariate models examined here is mathematical, and its ability to identify clinically consequential joint imbalance remains to be established. The article makes no recommendation for or against any particular implementation. DZ declares no competing interests. No other competing interests are declared.

\textbf{Use of generative AI.} The design-stage balance-scoring framework examined here originates in WJD's 2013 patent application~\cite{dwyer2013}, which predates this work. In preparing this manuscript WJD used a generative-AI assistant, Claude (Anthropic; Claude Opus 5.5 and earlier Claude models, accessed through the Claude app, September--November 2026), at the authors' direction, for: (i) literature searching and checking cited sources; (ii) drafting and editing prose from the authors' outlines, including first drafts of sections that the authors then revised; (iii) writing and checking the Python code (Additional files 2--5) that computes every derived number, runs the scheme comparison and draws every figure; (iv) writing up the derivations in Additional file 1 and checking them against Monte Carlo simulation; and (v) building and formatting tables. After the first submission, the assistant also suggested the simulated comparison of schemes and several of the added references; the authors reviewed both and adopted them. WJD also used ChatGPT (OpenAI; GPT-6 Astra, November 2026) to audit the revised manuscript and code; it proposed the proof of the two-sided boundary in Additional file 1 (D10), the prospectively fixed cut-off used for minimization in Additional file 4, and the local-separation calculations and two-covariate example in D13, and it identified an overstatement about multiple covariates in the conclusions of an earlier draft. Claude independently checked the D10 and D13 calculations and reran the corrected scheme comparison before the authors adopted them. The authors chose the research question and the 1/e threshold, selected the findings for inclusion, interpreted them, and take responsibility for the manuscript. The study uses no data other than published summary statistics, and all simulated data come from the deposited code with recorded seeds. Every derived number regenerates from that code (to Monte Carlo precision), every derivation is checked numerically, and every direct quotation and specific figure was checked against the full text of its source. The authors reviewed all AI-assisted content and take full responsibility for the work.

\textbf{Funding.} This research received no specific grant from any funding agency in the public, commercial, or not-for-profit sectors.

\textbf{Authors' contributions.} WJD conceived the study, conducted the literature review, derived the analytic results, wrote the software and simulations, and drafted the manuscript, with AI assistance as described under Use of generative AI. DZ reviewed the manuscript critically for statistical content, proposed the revisions to its structure, notation and presentation, including the small-trial check with t quantiles, and contributed to the interpretation of the results. Both authors read and approved the final manuscript.

\textbf{Acknowledgements.} Not applicable.

\hypertarget{additional-files}{%
\section{Additional files}\label{additional-files}}

In this preprint, Additional file 1 follows the reference list. Additional files 2--5 are included as arXiv ancillary files, with the expected output and the simulation results, and are archived on Zenodo.

\textbf{Additional file 1}. Derivations and supplementary tables and figures. Full derivations, with numerical checks, of every result marked \emph{derived} in the text (D1--D13); reference Tables A1--A9; Figures A1--A12; the simulated comparison of randomization schemes (Section N, Tables A10--A12).

\textbf{Additional file 2}. imbalance\_calculations.py. Python code that reproduces Additional file 1 Tables A3--A8, the worked examples and the Monte Carlo checks, including the small-trial check with t-tests.

\textbf{Additional file 3}. make\_figures.py. Python code that draws Figures 1--6 and Additional file 1 Figures A1--A12 from the formulas in Additional file 1 and from fixed-seed simulations.

\textbf{Additional file 4}. scheme\_comparison.py. Python code for the simulated comparison of randomization schemes: the check of the worst-of-k benchmark and the six-scheme comparison, with full results tabulated in Additional file 1, Section N.

\textbf{Additional file 5}. make\_figure7.py. Python code that draws Figure 7 from the output of Additional file 4.

\clearpage
\section{Additional file 1. Derivations and supplementary tables and figures}\label{additional-file-1}
\setcounter{figure}{0}\setcounter{table}{0}
\renewcommand{\thefigure}{A\arabic{figure}}\renewcommand{\thetable}{A\arabic{table}}
\hypertarget{scope-and-notation}{%
\subsection{Scope and notation}\label{scope-and-notation}}

This file gives full derivations of every result marked \emph{derived} in the main text, labelled D1--D13 to match the citations there. Each numerical result is reproduced by the Python script in Additional file 2 (\texttt{imbalance\_calculations.py}); its output is summarized in Section M. Figures A1--A12 are drawn by Additional file 3 (\texttt{make\_figures.py}). Tables A1--A9 are reference tables moved from the main text; Tables A3--A8 hold the numbers plotted in Figures 3--6 of the main text. Section N and Tables A10--A12 give the full results of the simulated comparison of randomization schemes (Additional file 4), plotted in Figure 7. The algebra in D1--D6 is standard omitted-variable and normal-theory algebra, written out here for completeness.

\textbf{Design.} Two arms with \(n\) patients each, \(N = 2n\), allocated by complete randomization with equal arm sizes. \(T_i = 1\) for treatment and \(0\) for control.

\textbf{Outcome model.}

\[Y_i = \mu + \tau T_i + \beta X_i + \varepsilon_i, \qquad \operatorname{E}(\varepsilon_i \mid X_i, T_i) = 0,\quad \operatorname{Var}(\varepsilon_i) = \sigma_\varepsilon^2. \qquad \text{(N1)}\]

The covariate \(X\) has variance \(\sigma_X^2\). Then

\[\sigma_Y^2 = \beta^2\sigma_X^2 + \sigma_\varepsilon^2, \qquad \rho = \operatorname{Corr}(X, Y \mid T) = \frac{\beta\sigma_X}{\sigma_Y}, \qquad \sigma_\varepsilon^2 = \sigma_Y^2(1-\rho^2). \qquad \text{(N2)}\]

\textbf{Summaries.} Arm means \(\bar x_T, \bar x_C, \bar y_T, \bar y_C\); \(\Delta_X = \bar x_T - \bar x_C\); standardized difference \(d = \Delta_X/\sigma_X\); unadjusted estimator \(\hat\tau = \bar y_T - \bar y_C\).

\textbf{Statistics.} With variances treated as known,

\[Z_{\text{bal}} = \frac{\Delta_X}{\sigma_X\sqrt{2/n}}, \qquad T_u = \frac{\hat\tau}{\sigma_Y\sqrt{2/n}}, \qquad P_{\text{bal}} = 2\{1-\Phi(|Z_{\text{bal}}|)\}. \qquad \text{(N3)}\]

\(\Phi\) and \(\phi\) are the standard normal distribution and density functions, and \(q = \Phi^{-1}(0.975) = 1.95996\).

\textbf{Sampling framework.} We treat patients as a random sample from a population and the allocation as independent of \((X, \varepsilon)\). Conditioning on the allocation \(A\) and on the realized covariate means is written \(\mid A\). The randomization-based (finite-population) framework, in which covariate values are fixed and only the allocation is random, gives the same first-order results; the normal-theory distributions in D6 and D10 are then large-sample approximations. Figure A1 shows how the quantities connect.

\begin{figure}
\centering
\includegraphics[width=1\textwidth,height=\textheight]{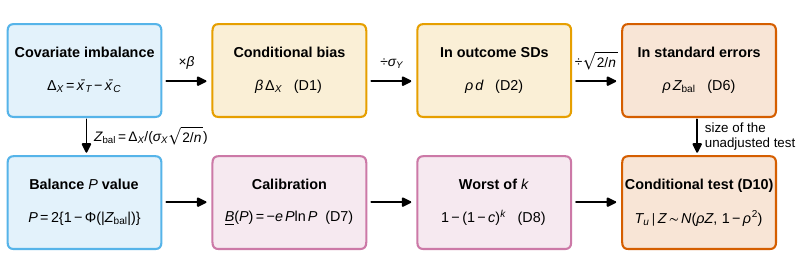}
\caption{Notation map. The top row follows a covariate imbalance to the conditional bias of the unadjusted estimate in outcome units and in standard errors; the bottom row follows the same imbalance through the balance statistic to the balance \emph{P} value, its calibration, the worst-of-k benchmark and the conditional size of the unadjusted test. Labels give the derivation in which each step is shown.}
\end{figure}

\hypertarget{d1.-unconditional-and-conditional-bias-of-the-unadjusted-estimator}{%
\subsection{D1. Unconditional and conditional bias of the unadjusted estimator}\label{d1.-unconditional-and-conditional-bias-of-the-unadjusted-estimator}}

Averaging (N1) within arms,

\[\bar y_T - \bar y_C = \tau + \beta(\bar x_T - \bar x_C) + (\bar\varepsilon_T - \bar\varepsilon_C). \qquad \text{(D1.1)}\]

Given the allocation and the covariate values, \(\operatorname{E}(\bar\varepsilon_T - \bar\varepsilon_C \mid A, X) = 0\), so

\[\operatorname{E}(\hat\tau \mid A, X) = \tau + \beta\Delta_X. \qquad \text{(D1.2)}\]

Under complete randomization with equal arms the arm labels are exchangeable, so \(\operatorname{E}_A(\Delta_X) = 0\) and, by iterated expectation,

\[\operatorname{E}(\hat\tau) = \tau + \beta\operatorname{E}_A(\Delta_X) = \tau. \qquad \text{(D1.3)}\]

The same holds for any restricted scheme that treats the two arms symmetrically, because symmetry again forces \(\operatorname{E}_A(\Delta_X) = 0\); this is the content of Morgan and Rubin's unbiasedness result for rerandomization~\cite{morgan2012}. The conditional bias \(\beta\Delta_X\) has mean zero over allocations and variance \(\beta^2\operatorname{Var}_A(\Delta_X)\); a scheme can change only the latter. \(\square\)

\begin{table}[tbp]
\centering\small
\caption{What proper randomization does and does not guarantee}
\begin{tabular}{@{}
  >{\raggedright\arraybackslash}p{(\columnwidth - 4\tabcolsep) * \real{0.3333}}
  >{\raggedright\arraybackslash}p{(\columnwidth - 4\tabcolsep) * \real{0.3333}}
  >{\raggedright\arraybackslash}p{(\columnwidth - 4\tabcolsep) * \real{0.3333}}@{}}
\toprule\noalign{}
\begin{minipage}[b]{\linewidth}\raggedright
Property
\end{minipage} & \begin{minipage}[b]{\linewidth}\raggedright
Guaranteed?
\end{minipage} & \begin{minipage}[b]{\linewidth}\raggedright
Note
\end{minipage} \\
\midrule\noalign{}

Unbiased unadjusted estimate, averaged over allocations & Yes & Complete and symmetric restricted schemes~\cite{morgan2012} \\
Protection from selection bias & Yes, if allocation is concealed and unpredictable & Predictable sequences weaken it~\cite{lachin1988a} \\
Valid unconditional \emph{P} values and confidence intervals & Yes & Chance imbalance is already allowed for~\cite{senn2013} \\
Balance on measured covariates in the realized trial & No & Large imbalances occur by chance~\cite{kahan2014,consort2025ee} \\
Unbiased estimate given the realized allocation & No, if a prognostic covariate is imbalanced and ignored & Removed by pre-specified adjustment~\cite{senn2013} \\
Balance on unmeasured covariates in the realized trial & No & Protected only in distribution \\
\bottomrule\noalign{}
\end{tabular}
\end{table}

\hypertarget{d2.-standardized-form-of-the-conditional-bias}{%
\subsection{D2. Standardized form of the conditional bias}\label{d2.-standardized-form-of-the-conditional-bias}}

Dividing (D1.2) minus \(\tau\) by \(\sigma_Y\) and using (N2),

\[\frac{\beta\Delta_X}{\sigma_Y} = \frac{\beta\sigma_X}{\sigma_Y}\cdot\frac{\Delta_X}{\sigma_X} = \rho\, d. \qquad \text{(D2.1)}\]

\emph{Example.} \(\rho = 0.5\), \(d = 0.2\) gives \(0.5 \times 0.2 = 0.10\) outcome SDs. \(\square\)

\hypertarget{d3.-change-scores-final-values-and-ancova}{%
\subsection{D3. Change scores, final values and ANCOVA}\label{d3.-change-scores-final-values-and-ancova}}

Suppose the baseline \(X\) and the outcome \(Y\) are the same measurement, so that \(Y - X\) is the change.

\textbf{Change score.} From (N1), \(Y - X = \mu + \tau T + (\beta - 1)X + \varepsilon\). Repeating D1,

\[\operatorname{E}\{(\bar y_T - \bar x_T) - (\bar y_C - \bar x_C) \mid A, X\} = \tau + (\beta - 1)\Delta_X. \qquad \text{(D3.1)}\]

If \(\sigma_X = \sigma_Y\), then \(\beta = \rho\), and for \(0 < \rho < 1\) the coefficient \(\beta - 1\) is negative while \(\beta\) is positive. The final-value analysis (coefficient \(\beta\)) and the change-score analysis (coefficient \(\beta - 1\)) are therefore biased in opposite directions by the same imbalance, which is the pattern shown in Figure 1 of the main text and summarized in Table A2. The two biases have equal magnitude when \(\beta = 1/2\).

\textbf{ANCOVA.} The least-squares fit of \(Y\) on \((1, T, X)\) gives

\[\hat\tau_{\text{ANCOVA}} = (\bar y_T - \bar y_C) - \hat\beta(\bar x_T - \bar x_C), \qquad \text{(D3.2)}\]

where \(\hat\beta\) is the pooled within-arm slope. Conditional on \(A\) and \(X\), ordinary least squares is unbiased for \((\mu, \tau, \beta)\) under (N1), so \(\operatorname{E}(\hat\tau_{\text{ANCOVA}} \mid A, X) = \tau\) for every value of \(\Delta_X\).

\textbf{Precision.} Conditional on \(A\) and \(X\),

\[\operatorname{Var}(\hat\tau_{\text{ANCOVA}} \mid A, X) = \sigma_\varepsilon^2\left\{\frac{2}{n} + \frac{\Delta_X^2}{S_{XX}}\right\}, \qquad \text{(D3.3)}\]

with \(S_{XX}\) the pooled within-arm sum of squares of \(X\). Since \(\Delta_X^2/S_{XX} = O_p(1/n^2)\), smaller than \(2/n\) by a factor of order \(1/n\), to first order

\[\frac{\operatorname{Var}(\hat\tau_{\text{ANCOVA}})}{\operatorname{Var}(\hat\tau)} \approx \frac{\sigma_\varepsilon^2}{\sigma_Y^2} = 1 - \rho^2, \qquad \frac{\operatorname{SE}(\hat\tau_{\text{ANCOVA}})}{\operatorname{SE}(\hat\tau)} \approx \sqrt{1 - \rho^2}. \qquad \text{(D3.4)}\]

At \(\rho = 0.9\) this ratio is \(0.44\).

\textbf{Exact variance inflation.} Write (D3.3) as \(\operatorname{Var}(\hat\tau_{\text{ANCOVA}} \mid A, X) = (2\sigma_\varepsilon^2/n)\,\text{VIF}\), with \(\text{VIF} = 1 + n\Delta_X^2/(2S_{XX})\). With \(s_X^2 = S_{XX}/(2n - 2)\), the pooled within-arm variance of \(X\), and \(t_{\text{bal}} = \Delta_X/(s_X\sqrt{2/n})\), the pooled-variance balance \(t\) statistic,

\[\text{VIF} = 1 + \frac{t_{\text{bal}}^2}{2n - 2}. \qquad \text{(D3.5)}\]

This is the variance inflation factor of Senn, König and Posch~\cite{senn2026}. For a normal covariate under complete randomization \(t_{\text{bal}}\) has a \(t\) distribution with \(2n - 2\) degrees of freedom, so \(\operatorname{E}(t_{\text{bal}}^2) = (2n - 2)/(2n - 4)\) and \(\operatorname{E}(\text{VIF}) = 1 + 1/(2n - 4)\), the expectation they attribute to Cochran. When the balance \(t\) test gives \emph{P} = 1/e, VIF = 1.022 with \(n = 20\) per arm and 1.004 with \(n = 100\); at \emph{P} = 0.05 it is 1.108 and 1.020. The inflation is a matter of precision only: the estimate stays conditionally unbiased. \(\square\)

\begin{table}[tbp]
\centering\small
\caption{Analysis choices and their effect on imbalance bias}
\begin{tabular}{@{}
  >{\raggedright\arraybackslash}p{(\columnwidth - 6\tabcolsep) * \real{0.2500}}
  >{\raggedright\arraybackslash}p{(\columnwidth - 6\tabcolsep) * \real{0.2500}}
  >{\raggedright\arraybackslash}p{(\columnwidth - 6\tabcolsep) * \real{0.2500}}
  >{\raggedright\arraybackslash}p{(\columnwidth - 6\tabcolsep) * \real{0.2500}}@{}}
\toprule\noalign{}
\begin{minipage}[b]{\linewidth}\raggedright
Choice
\end{minipage} & \begin{minipage}[b]{\linewidth}\raggedright
Conditional bias
\end{minipage} & \begin{minipage}[b]{\linewidth}\raggedright
Precision and error rates
\end{minipage} & \begin{minipage}[b]{\linewidth}\raggedright
Sources
\end{minipage} \\
\midrule\noalign{}

Final value, no covariates & βΔ, same sign as the imbalance & Valid unconditionally; miscalibrated conditionally & ~\cite{egbewale2014,vickers2001} \\
Change score & (β − 1)Δ, sign reversed & Efficient only at high correlation & ~\cite{egbewale2014,vickers2001} \\
Pre-specified ANCOVA on prognostic covariates & Removed & Most precise; large power gains & ~\cite{kahan2014,egbewale2014} \\
Omitting stratification or minimization factors & Not a bias issue & SEs too large; type I error and power too low & ~\cite{kahan2012sim,fda2023} \\
Omitting centre in a centre-stratified trial & Not a bias issue & SE inflated by √\{1/(1 − ICC)\} & ~\cite{kahan2013centre} \\
Post hoc adjustment chosen because of imbalance & Removed, but data-driven & Credible only as a sensitivity analysis & ~\cite{ema2015,iche9} \\
Unadjusted odds or hazard ratio & Adds non-collapsibility & Attenuated estimate, less power & ~\cite{gail1984,hernandez2004} \\
\bottomrule\noalign{}
\end{tabular}
\end{table}

\begin{figure}
\centering
\includegraphics[width=1\textwidth,height=\textheight]{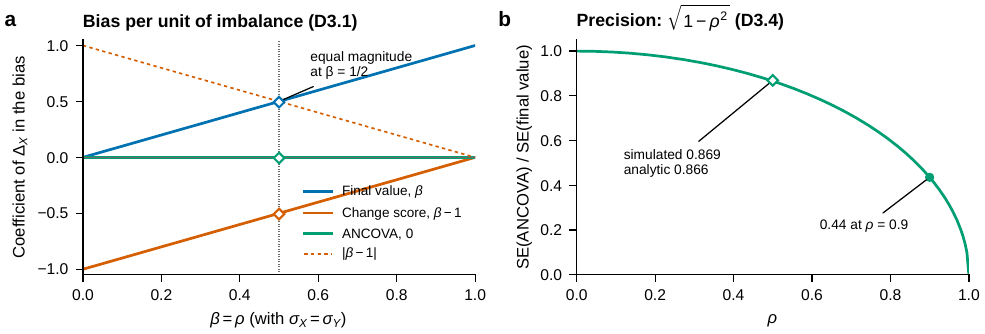}
\caption{Change scores, final values and ANCOVA (D3), with \(\sigma_X = \sigma_Y\) so that \(\beta = \rho\). (a) Coefficient of \(\Delta_X\) in the conditional bias: \(\beta\) for the final value, \(\beta - 1\) for the change score and 0 for ANCOVA; the dotted line is \(|\beta - 1|\), which equals \(\beta\) at \(\beta = 1/2\). (b) Ratio of the standard error of the ANCOVA estimate to that of the final-value estimate, \(\sqrt{1-\rho^2}\) (D3.4). Diamonds are fitted values from 20 000 simulated trials with 100 patients per arm and \(\rho = 0.5\) (Additional file 3; the seeds differ from Additional file 2, so the third decimal differs from Section M).}
\end{figure}

\hypertarget{d4.-binary-outcome-with-a-high-risk-subgroup}{%
\subsection{D4. Binary outcome with a high-risk subgroup}\label{d4.-binary-outcome-with-a-high-risk-subgroup}}

Let \(H \in \{0, 1\}\) mark high-risk patients, with control-arm event risks \(r_H\) and \(r_L\), and suppose the treatment changes the risk by the same amount \(\delta\) in both strata (constant risk difference). Let \(\pi_{H,T}\) and \(\pi_{H,C}\) be the realized proportions of high-risk patients in the two arms. Conditional on these proportions,

\[\operatorname{E}(\hat p_T) = \pi_{H,T}(r_H + \delta) + (1 - \pi_{H,T})(r_L + \delta) = r_L + \delta + \pi_{H,T}(r_H - r_L), \qquad \text{(D4.1)}\]

\[\operatorname{E}(\hat p_C) = r_L + \pi_{H,C}(r_H - r_L). \qquad \text{(D4.2)}\]

Subtracting,

\[\operatorname{E}(\hat p_T - \hat p_C \mid \pi_{H,T}, \pi_{H,C}) = \delta + (\pi_{H,T} - \pi_{H,C})(r_H - r_L). \qquad \text{(D4.3)}\]

This is equation (5) of the main text. It is D1.2 with \(X = H\) and \(\beta = r_H - r_L\), because a linear probability model is exact for a binary covariate when the risk difference is constant.

\emph{Examples.} (i) \((\pi_{H,T} - \pi_{H,C}) = 0.10\) and \(r_H - r_L = 0.30\) give a bias of \(0.030\). (ii) With survival to discharge or 1 year among actively treated infants of \(55.8\%\) at 23 weeks and \(94.7\%\) at 28 weeks~\cite{bell2022}, the mortality difference between the strata is \(0.947 - 0.558 = 0.389\). A 10-point excess of 23-week infants in one arm gives a bias of \(0.10 \times 0.389 = 0.039\), about 3.9 percentage points in the mortality difference. This is illustrative arithmetic: it treats a mixture of only those two gestational ages and assumes a constant risk difference. \(\square\)

\hypertarget{d5.-frequency-of-chance-imbalance}{%
\subsection{D5. Frequency of chance imbalance}\label{d5.-frequency-of-chance-imbalance}}

\textbf{At least one ``significant'' covariate.} For \(k\) independent balance tests, each with a uniform \emph{P} value under complete randomization,

\[\Pr(\text{at least one } P < 0.05) = 1 - 0.95^k. \qquad \text{(D5.1)}\]

This gives \(0.05, 0.226, 0.401, 0.537, 0.642, 0.785\) for \(k = 1, 5, 10, 15, 20, 30\) (Table A3). For the trial of de Boer and colleagues~\cite{deboer2015}, the probability of exactly one significant result among 12 independent true nulls is

\[\binom{12}{1}(0.05)(0.95)^{11} = 0.341. \qquad \text{(D5.2)}\]

\begin{table}[tbp]
\centering\small
\caption{Probability of at least one baseline \emph{P} \textless{} 0.05 among k independent covariates (any sample size)}
\begin{tabular}{@{}
  >{\raggedright\arraybackslash}p{(\columnwidth - 12\tabcolsep) * \real{0.1429}}
  >{\raggedright\arraybackslash}p{(\columnwidth - 12\tabcolsep) * \real{0.1429}}
  >{\raggedright\arraybackslash}p{(\columnwidth - 12\tabcolsep) * \real{0.1429}}
  >{\raggedright\arraybackslash}p{(\columnwidth - 12\tabcolsep) * \real{0.1429}}
  >{\raggedright\arraybackslash}p{(\columnwidth - 12\tabcolsep) * \real{0.1429}}
  >{\raggedright\arraybackslash}p{(\columnwidth - 12\tabcolsep) * \real{0.1429}}
  >{\raggedright\arraybackslash}p{(\columnwidth - 12\tabcolsep) * \real{0.1429}}@{}}
\toprule\noalign{}
\begin{minipage}[b]{\linewidth}\raggedright
k
\end{minipage} & \begin{minipage}[b]{\linewidth}\raggedright
1
\end{minipage} & \begin{minipage}[b]{\linewidth}\raggedright
5
\end{minipage} & \begin{minipage}[b]{\linewidth}\raggedright
10
\end{minipage} & \begin{minipage}[b]{\linewidth}\raggedright
15
\end{minipage} & \begin{minipage}[b]{\linewidth}\raggedright
20
\end{minipage} & \begin{minipage}[b]{\linewidth}\raggedright
30
\end{minipage} \\
\midrule\noalign{}

Pr(at least one \emph{P} \textless{} 0.05) & 0.05 & 0.23 & 0.40 & 0.54 & 0.64 & 0.79 \\
\bottomrule\noalign{}
\end{tabular}
\end{table}

\textbf{Size of the standardized difference.} Under complete randomization with \(n\) per arm, \(\operatorname{Var}(\bar x_T - \bar x_C) = 2\sigma_X^2/n\), so \(d \sim N(0, 2/n)\) exactly for normal \(X\) and approximately otherwise. Hence

\[\Pr(|d| > c) = 2\,\Phi\!\left(-\frac{c}{\sqrt{2/n}}\right) = 2\,\Phi\!\left(-c\sqrt{n/2}\right). \qquad \text{(D5.3)}\]

For \(c = 0.2\) and \(n = 25\), \(c\sqrt{n/2} = 0.707\) and the probability is \(0.480\); for \(n = 200\), \(c\sqrt{n/2} = 2\) and it is \(0.046\). The full set of values is Table A4. \(\square\)

\begin{table}[tbp]
\centering\small
\caption{Probability that the standardized difference \textbar d\textbar{} exceeds a threshold under complete randomization}
\begin{tabular}{@{}lllllll@{}}
\toprule\noalign{}
Patients per arm & 25 & 50 & 100 & 200 & 500 & 1000 \\
\midrule\noalign{}

Pr(\textbar d\textbar{} \textgreater{} 0.1) & 0.72 & 0.62 & 0.48 & 0.32 & 0.11 & 0.025 \\
Pr(\textbar d\textbar{} \textgreater{} 0.2) & 0.48 & 0.32 & 0.16 & 0.046 & 0.002 & \textless0.001 \\
\bottomrule\noalign{}
\end{tabular}
\end{table}

\begin{figure}
\centering
\includegraphics[width=0.62\textwidth,height=\textheight]{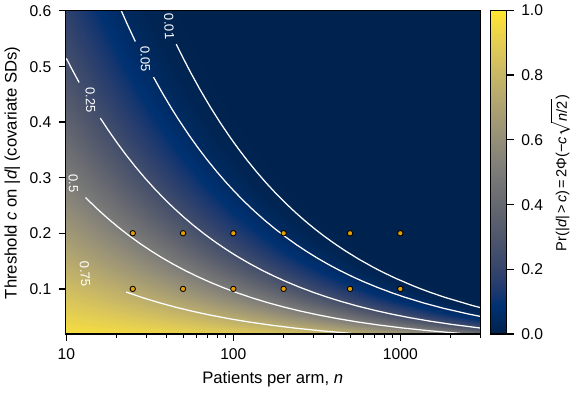}
\caption{Probability that the standardized difference exceeds a threshold, \(\Pr(|d| > c) = 2\Phi(-c\sqrt{n/2})\) (D5.3), over patients per arm \(n\) and threshold \(c\). White contours at 0.01, 0.05, 0.25, 0.5 and 0.75; points mark the entries of Table A4 (\(c = 0.1\) and \(0.2\)).}
\end{figure}

\hypertarget{d6.-bias-in-standard-error-units-does-not-shrink-with-sample-size}{%
\subsection{D6. Bias in standard-error units does not shrink with sample size}\label{d6.-bias-in-standard-error-units-does-not-shrink-with-sample-size}}

The standard error of the unadjusted estimator is \(\operatorname{SE}(\hat\tau) = \sigma_Y\sqrt{2/n}\). Dividing the conditional bias by it, and using (N3),

\[\frac{\beta\Delta_X}{\sigma_Y\sqrt{2/n}} = \frac{\beta\sigma_X}{\sigma_Y}\cdot\frac{\Delta_X}{\sigma_X\sqrt{2/n}} = \rho\,Z_{\text{bal}}. \qquad \text{(D6.1)}\]

Under complete randomization \(Z_{\text{bal}} \sim N(0, 1)\) for every \(n\) (exactly for normal \(X\), asymptotically otherwise). The conditional bias in SE units therefore has the distribution \(N(0, \rho^2)\) at every sample size. In outcome units, by contrast, the bias is \(\rho\sigma_Y d\) with \(\operatorname{Var}(d) = 2/n\), which shrinks with \(n\). Both statements are true; inference (type I error, coverage) depends on the first. Senn~\cite{senn1989} made the same point for the conditional size of the test of efficacy, which does not depend on \(n\). \(\square\)

\hypertarget{d7.-sellkebayarriberger-calibration}{%
\subsection{D7. Sellke--Bayarri--Berger calibration}\label{d7.-sellkebayarriberger-calibration}}

Sellke, Bayarri and Berger~\cite{sellke2001} show that, for a precise null hypothesis and a broad class of alternatives under which the \emph{P} value has a decreasing density,

\[B(P) \ge \underline{B}(P) = \begin{cases} -e\,P\ln P, & P < 1/e,\\ 1, & P \ge 1/e,\end{cases} \qquad \text{(D7.1)}\]

where \(B(P)\) is the Bayes factor of the null to the alternative. The corresponding lower bound on the conditional (posterior, with equal prior odds) error probability is

\[\alpha(P) = \left\{1 + \frac{1}{\underline{B}(P)}\right\}^{-1}. \qquad \text{(D7.2)}\]

Benjamin and Berger~\cite{benjamin2019} recommend reporting \(1/\underline{B}(P)\) as the maximum data-based odds for the alternative, crediting Vovk~\cite{vovk1993} and Sellke and colleagues.

\textbf{Why the bound reaches 1 at \(1/e\).} Let \(g(P) = -e\,P\ln P\). Then \(g'(P) = -e(\ln P + 1)\), which is zero at \(P = 1/e\), and \(g(1/e) = -e \cdot e^{-1} \cdot (-1) = 1\). So \(g\) rises to its maximum value of 1 exactly at \(P = 1/e\), and the bound is continuous there.

\emph{Worked values.} At \(P = 0.05\): \(\ln 0.05 = -2.9957\), \(\underline{B} = e \times 0.05 \times 2.9957 = 0.4072\), \(1/\underline{B} = 2.46\), \(\alpha = 1/(1 + 2.456) = 0.289\). These match the values printed by Sellke and colleagues. The remaining rows of Table A5 follow in the same way.

\begin{table}[tbp]
\centering\small
\caption{Sellke--Bayarri--Berger calibration (derived from the published formula)}
\begin{tabular}{@{}
  >{\raggedright\arraybackslash}p{(\columnwidth - 6\tabcolsep) * \real{0.2500}}
  >{\raggedright\arraybackslash}p{(\columnwidth - 6\tabcolsep) * \real{0.2500}}
  >{\raggedright\arraybackslash}p{(\columnwidth - 6\tabcolsep) * \real{0.2500}}
  >{\raggedright\arraybackslash}p{(\columnwidth - 6\tabcolsep) * \real{0.2500}}@{}}
\toprule\noalign{}
\begin{minipage}[b]{\linewidth}\raggedright
\emph{P}
\end{minipage} & \begin{minipage}[b]{\linewidth}\raggedright
Lower bound on Bayes factor B(\emph{P})
\end{minipage} & \begin{minipage}[b]{\linewidth}\raggedright
Maximum odds for the alternative, 1/B(\emph{P})
\end{minipage} & \begin{minipage}[b]{\linewidth}\raggedright
α(\emph{P})
\end{minipage} \\
\midrule\noalign{}

0.10 & 0.626 & 1.60 & 0.385 \\
0.05 & 0.407 & 2.46 & 0.289 \\
0.01 & 0.125 & 7.99 & 0.111 \\
0.005 & 0.072 & 13.9 & 0.067 \\
0.001 & 0.0188 & 53.3 & 0.018 \\
1/e ≈ 0.368 & 1.000 & 1.00 & 0.500 \\
\bottomrule\noalign{}
\end{tabular}
\end{table}

\textbf{The same point in Fisher's combination.} Under the null \(P \sim\) Uniform\((0,1)\), so \(-2\ln P \sim \chi^2_2\) with mean 2. The contribution \(-2\ln P\) equals its null expectation exactly when \(P = e^{-1}\): a \emph{P} value below \(1/e\) pushes Fisher's combined statistic above its null mean and a \emph{P} value above \(1/e\) pulls it below. Replicating one \emph{P} value \(n\) times makes the point exact in the limit: \(-2n\ln P\) is compared with \(\chi^2_{2n}\), whose mean is \(2n\) and whose standard deviation is \(2\sqrt{n}\), so the combined \emph{P} value tends to 0 if \(P < e^{-1}\) and to 1 if \(P > e^{-1}\). Elston~\cite{elston1991} made this argument with the median of \(\chi^2_{2n}\), which is close to its mean for large \(n\), setting \(-2n\ln P = 2n\) to obtain \(P = e^{-1} \approx 0.37\); he also showed that Fisher's method follows from a weight-free ranking of \emph{P} value vectors. Fisher~\cite{fisher1948} described the same mechanism without the number: a \emph{P} value above one half ``will tend to lower'' the significance of the combined test because it introduces ``two additional degrees of freedom''. \(\square\)

\begin{figure}
\centering
\includegraphics[width=0.62\textwidth,height=\textheight]{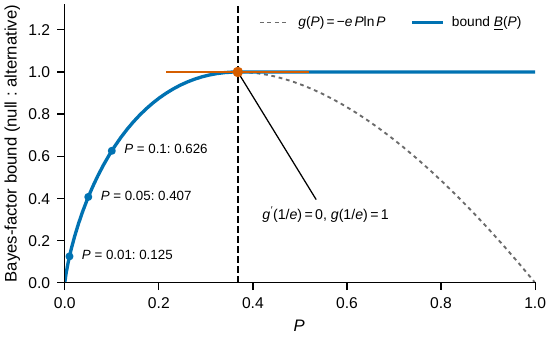}
\caption{Why the Sellke--Bayarri--Berger bound reaches 1 at \(1/e\) (D7.1). The dashed grey curve is \(g(P) = -e\,P\ln P\); the blue curve is the bound, equal to \(g\) below \(1/e\) and to 1 above it. At \(P = 1/e\) the derivative of \(g\) is zero and \(g = 1\), so the bound is continuous. Points mark \(P = 0.01\), 0.05 and 0.10.}
\end{figure}

\hypertarget{d8.-worst-of-k-benchmark-under-complete-randomization}{%
\subsection{D8. Worst-of-k benchmark under complete randomization}\label{d8.-worst-of-k-benchmark-under-complete-randomization}}

\textbf{Uniformity.} For a continuous balance statistic with a continuous null distribution, the \emph{P} value is Uniform\((0,1)\) under complete randomization, because \(\Pr(P \le u) = \Pr\{|Z_{\text{bal}}| \ge \Phi^{-1}(1 - u/2)\} = u\). This holds for every \(n\) and for every value of \(\rho\), because the balance statistic does not involve the outcome.

\textbf{Worst of \(k\).} For \(k\) independent covariates, the worst (smallest) \emph{P} value \(P_{(1)}\) satisfies

\[\Pr(P_{(1)} \le c) = 1 - \Pr(\text{all } P_j > c) = 1 - (1 - c)^k. \qquad \text{(D8.1)}\]

With \(c = 1/e\), \(1 - c = 0.63212\), giving \(0.368, 0.600, 0.747, 0.899, 0.990, 0.999\) for \(k = 1, 2, 3, 5, 10, 15\) (Table A6). The probability that all ten \emph{P} values are at least \(1/e\) is \(0.63212^{10} = 0.0102\).

\begin{table}[tbp]
\centering\small
\caption{Probability that the worst balance \emph{P} value is at or below a threshold under complete randomization, k independent covariates}
\begin{tabular}{@{}lllllll@{}}
\toprule\noalign{}
k & 1 & 2 & 3 & 5 & 10 & 15 \\
\midrule\noalign{}

Pr(worst \emph{P} ≤ 1/e) & 0.37 & 0.60 & 0.75 & 0.90 & 0.99 & \textgreater0.99 \\
Pr(worst \emph{P} ≤ 0.05) & 0.05 & 0.10 & 0.14 & 0.23 & 0.40 & 0.54 \\
\bottomrule\noalign{}
\end{tabular}
\end{table}

\textbf{Dependence and discreteness.} For jointly normal balance statistics with any correlation structure, Šidák's inequality for two-sided tests gives \(\Pr(\text{all } |Z_j| \le c_j) \ge \prod_j \Pr(|Z_j| \le c_j)\), so the worst \emph{P} value is stochastically larger than under independence and (D8.1) is an upper bound on the benchmark flag rate. Exact tests on categorical covariates give discrete, typically conservative \emph{P} values, so the benchmark for a mixed battery should be obtained by simulation. Figure A5b shows the size of the effect for ten equicorrelated normal balance statistics: the flag rate \(\Pr(P_{(1)} \le 1/e)\) falls from 0.990 under independence to 0.96 at a correlation of 0.5 and 0.67 at 0.9 (simulation, 200 000 draws per point; Additional file 3). \(\square\)

\begin{figure}
\centering
\includegraphics[width=1\textwidth,height=\textheight]{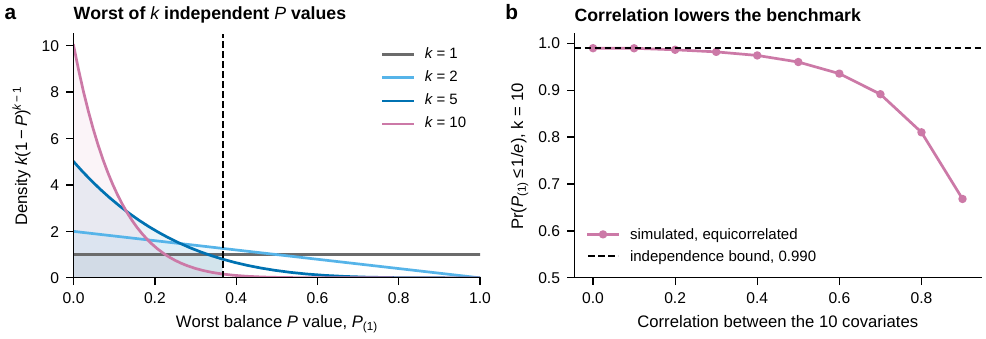}
\caption{The worst-of-\(k\) benchmark (D8). (a) Density of the smallest of \(k\) independent uniform \emph{P} values, \(k(1-P)^{k-1}\), for \(k = 1, 2, 5\) and 10; the shaded area to the left of the dashed line (\(P = 1/e\)) is the benchmark flag rate \(1 - (1-1/e)^k\). (b) Flag rate for \(k = 10\) when the ten balance statistics are equicorrelated normal variables, by simulation (200 000 draws per point); the dashed line is the independence value, 0.990, which is an upper bound for any correlation (Šidák's inequality).}
\end{figure}

\hypertarget{d9.-bias-in-standard-errors-at-a-given-balance-p-value}{%
\subsection{\texorpdfstring{D9. Bias in standard errors at a given balance \emph{P} value}{D9. Bias in standard errors at a given balance P value}}\label{d9.-bias-in-standard-errors-at-a-given-balance-p-value}}

From (D6.1), with \(|Z_{\text{bal}}| = \Phi^{-1}(1 - P/2)\),

\[\frac{|\text{bias}|}{\operatorname{SE}(\hat\tau)} = \rho\,\Phi^{-1}(1 - P/2). \qquad \text{(D9.1)}\]

At \(P = 1/e\), \(\Phi^{-1}(1 - 0.18394) = \Phi^{-1}(0.81606) = 0.9005\), so the bias is \(0.27, 0.45, 0.63, 0.81\) SE for \(\rho = 0.3, 0.5, 0.7, 0.9\). At \(P = 0.05\) the multiplier is \(1.960\). The full set of values is Table A7. \(\square\)

\begin{table}[tbp]
\centering\small
\caption{Conditional bias of the unadjusted estimate, in standard errors, at a given balance \emph{P} value}
\begin{tabular}{@{}llllll@{}}
\toprule\noalign{}
Balance \emph{P} & \textbar Z\textsubscript{bal}\textbar{} & ρ = 0.3 & ρ = 0.5 & ρ = 0.7 & ρ = 0.9 \\
\midrule\noalign{}

0.50 & 0.67 & 0.20 & 0.34 & 0.47 & 0.61 \\
1/e ≈ 0.368 & 0.90 & 0.27 & 0.45 & 0.63 & 0.81 \\
0.20 & 1.28 & 0.38 & 0.64 & 0.90 & 1.15 \\
0.05 & 1.96 & 0.59 & 0.98 & 1.37 & 1.76 \\
0.01 & 2.58 & 0.77 & 1.29 & 1.80 & 2.32 \\
\bottomrule\noalign{}
\end{tabular}
\end{table}

\hypertarget{d10.-conditional-type-i-error-of-the-unadjusted-test}{%
\subsection{D10. Conditional type I error of the unadjusted test}\label{d10.-conditional-type-i-error-of-the-unadjusted-test}}

\textbf{Joint distribution.} Under \(H_0: \tau = 0\), (D1.1) gives

\[T_u = \frac{\beta\Delta_X + (\bar\varepsilon_T - \bar\varepsilon_C)}{\sigma_Y\sqrt{2/n}} = \rho\,Z_{\text{bal}} + \sqrt{1 - \rho^2}\;W, \qquad W = \frac{\bar\varepsilon_T - \bar\varepsilon_C}{\sigma_\varepsilon\sqrt{2/n}}, \qquad \text{(D10.1)}\]

using \(\beta\sigma_X/\sigma_Y = \rho\) and \(\sigma_\varepsilon/\sigma_Y = \sqrt{1 - \rho^2}\). For normal data \(W \sim N(0,1)\) independently of \(Z_{\text{bal}}\), so \((Z_{\text{bal}}, T_u)\) is bivariate standard normal with correlation \(\rho\), and

\[T_u \mid Z_{\text{bal}} = z \;\sim\; N\!\left(\rho z,\; 1 - \rho^2\right). \qquad \text{(D10.2)}\]

This is the model of Senn~\cite{senn1989,senn1994}, with known dispersion parameters. In 1989 he treated the one-sided test; (D10.5) below is his equation (3) in the present notation, and in 1993 he applied it to crude outcomes, change from baseline and ANCOVA~\cite{senn1993}. Permutt wrote the two-sided conditional rejection probability of the unadjusted test, (D10.3) below, in the appendix of a 1990 paper, noting that it is even and increasing in \(|z|\) and averages to the nominal level~\cite{permutt1990}. In 1994 Senn gave (D10.3) as his equation (2), for arms of sizes \(n_1\) and \(n_2\), with the worked value 0.09 for \(\rho = 0.73\) and \(z = 1.43\); (D10.3) gives 0.0901. The expansion (D10.4), the boundary \(|z| \le 1\) that holds for every \(\rho\), and the boundary table are not in any of these papers.

\textbf{Two-sided conditional size.} Write \(m = \rho z\) and \(s = \sqrt{1 - \rho^2}\). Then

\[\alpha_2(z, \rho) = \Pr(|T_u| > q \mid z) = \Phi\!\left(\frac{-q - m}{s}\right) + \Phi\!\left(\frac{m - q}{s}\right). \qquad \text{(D10.3)}\]

\emph{Monotonicity in \(|z|\).} \(\partial\alpha_2/\partial m = s^{-1}\{\phi((m - q)/s) - \phi((m + q)/s)\}\), which is positive for \(m > 0\) because \(|m - q| < m + q\). So for fixed \(\rho\) the conditional size increases with the imbalance, and for \(0 < |\rho| < 1\) there is a unique boundary \(z^*(\rho)\) at which \(\alpha_2 = 0.05\) (at \(\rho = 0\) the size is 0.05 for every \(z\)).

\emph{Small-\(\rho\) expansion.} Expanding (D10.3) in \(\rho\) (verified symbolically with SymPy) gives

\[\alpha_2(z, \rho) = \alpha + q\,\phi(q)\left[\rho^2(z^2 - 1) + \frac{(q^2 - 3)(z^4 - 6z^2 + 3)}{12}\,\rho^4\right] + O(\rho^6), \qquad \text{(D10.4)}\]

with \(\alpha = 0.05\). The leading term comes from two competing effects: the shift \(m^2 q\phi(q)\) from the conditional mean, and the reduction \(-\rho^2 q\phi(q)\) from the smaller conditional variance. To second order, the size exceeds nominal exactly when \(|z| > 1\), that is, when \(P_{\text{bal}} < 2\Phi(-1) \approx 0.3173\). At \(|z| = 1\) the second-order term vanishes and the fourth-order term is \(-q\phi(q)(q^2 - 3)\rho^4/6 < 0\), since \(q^2 = 3.84 > 3\). So the size at \(|z| = 1\) is below nominal for small \(\rho > 0\).

\emph{The boundary that holds for every \(\rho\).} \textbf{Proposition.} Under (D10.2), \(\alpha_2(z, \rho) \le 0.05\) for every \(\rho \in (0, 1)\) whenever \(|z| \le 1\), that is, whenever \(P_{\text{bal}} \ge 2\Phi(-1) = 0.317310\ldots\); and for every \(|z| > 1\) there is a \(\rho > 0\) with \(\alpha_2(z, \rho) > 0.05\). The region \(|z| \le 1\) is therefore the sharp region of two-sided error control that does not depend on \(\rho\).

\emph{Proof.} By the monotonicity in \(|z|\) it suffices to take \(z = 1\). Write \(r = \rho\), \(u = (q - r)/s\) and \(v = (q + r)/s\), so that \(\alpha_2(1, r) = \Phi(-u) + \Phi(-v)\). Differentiating,

\[s^3\,\frac{\partial \alpha_2(1, r)}{\partial r} = (1 - qr)\,\phi(u) - (1 + qr)\,\phi(v). \qquad \text{(D10.4a)}\]

If \(qr \ge 1\) both terms are non-positive and the second is negative. If \(0 < qr < 1\), then \(\phi(u)/\phi(v) = \exp\{2qr/(1 - r^2)\}\), and the derivative is negative exactly when \(H(r) > 0\), where

\[H(r) = \operatorname{artanh}(qr) - \frac{qr}{1 - r^2}, \qquad H'(r) = \frac{q r^2\{(q^2 - 3) + (q^2 + 1) r^2\}}{(1 - q^2 r^2)(1 - r^2)^2}. \qquad \text{(D10.4b)}\]

Since \(H(0) = 0\) and \(q^2 = 3.84 > 3\), \(H'(r) > 0\) and hence \(H(r) > 0\) on \((0, 1/q)\). So \(\alpha_2(1, r)\) decreases on \((0, 1)\) from its limit \(\alpha_2(1, 0) = 0.05\), and \(\alpha_2(z, \rho) < 0.05\) for \(|z| \le 1\) and every \(\rho \in (0, 1)\). Conversely, for \(|z| > 1\), (D10.4) gives \(\alpha_2(z, \rho) - 0.05 = q\phi(q)(z^2 - 1)\rho^2 + O(\rho^4) > 0\) for small \(\rho > 0\). \(\square\)

Because the region is sharp as \(\rho \to 0\), a rounded cut-off below \(2\Phi(-1)\) does not guarantee control: at \(P_{\text{bal}} = 0.317\) and \(\rho = 0.05\), (D10.3) gives 0.0500003. The bound is therefore stated as \(2\Phi(-1)\) throughout. Since \(1/e = 0.368 > 2\Phi(-1)\), equivalently \(|z| = 0.90 < 1\) at \(P_{\text{bal}} = 1/e\), the 1/e threshold lies inside the region for every \(\rho\). The result concerns one normal covariate with known variances; the small-trial check with t-tests (Section M) and the multi-covariate simulation (Section N) are separate.

\emph{Crossings at fixed \(\rho\).} For a given \(\rho\), the crossing \(z^*(\rho)\) at which \(\alpha_2 = 0.05\), found numerically, moves to larger \(|z|\) (smaller \(P_{\text{bal}}\)) as \(\rho\) grows:

\begin{center}\small
\begin{tabular}{@{}
  >{\raggedright\arraybackslash}p{(\columnwidth - 14\tabcolsep) * \real{0.1250}}
  >{\raggedright\arraybackslash}p{(\columnwidth - 14\tabcolsep) * \real{0.1250}}
  >{\raggedright\arraybackslash}p{(\columnwidth - 14\tabcolsep) * \real{0.1250}}
  >{\raggedright\arraybackslash}p{(\columnwidth - 14\tabcolsep) * \real{0.1250}}
  >{\raggedright\arraybackslash}p{(\columnwidth - 14\tabcolsep) * \real{0.1250}}
  >{\raggedright\arraybackslash}p{(\columnwidth - 14\tabcolsep) * \real{0.1250}}
  >{\raggedright\arraybackslash}p{(\columnwidth - 14\tabcolsep) * \real{0.1250}}
  >{\raggedright\arraybackslash}p{(\columnwidth - 14\tabcolsep) * \real{0.1250}}@{}}
\toprule\noalign{}
\begin{minipage}[b]{\linewidth}\raggedright
\(\rho\)
\end{minipage} & \begin{minipage}[b]{\linewidth}\raggedright
0.01
\end{minipage} & \begin{minipage}[b]{\linewidth}\raggedright
0.10
\end{minipage} & \begin{minipage}[b]{\linewidth}\raggedright
0.30
\end{minipage} & \begin{minipage}[b]{\linewidth}\raggedright
0.50
\end{minipage} & \begin{minipage}[b]{\linewidth}\raggedright
0.70
\end{minipage} & \begin{minipage}[b]{\linewidth}\raggedright
0.90
\end{minipage} & \begin{minipage}[b]{\linewidth}\raggedright
0.95
\end{minipage} \\
\midrule\noalign{}
\(z^*(\rho)\) & 1.0000 & 1.0007 & 1.0082 & 1.0348 & 1.1213 & 1.3811 & 1.5225 \\
boundary \(P_{\text{bal}}\) & 0.3173 & 0.3170 & 0.3134 & 0.3007 & 0.2622 & 0.1672 & 0.1279 \\
\bottomrule\noalign{}
\end{tabular}
\end{center}

These fixed-\(\rho\) values answer a different question from the proposition: they apply only when \(\rho\) is known.

\begin{figure}
\centering
\includegraphics[width=1\textwidth,height=\textheight]{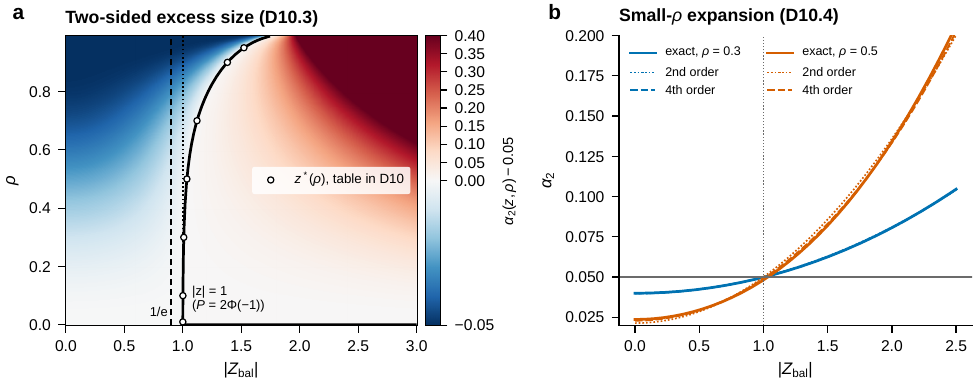}
\caption{Two-sided conditional size (D10.3). (a) Excess size \(\alpha_2(z,\rho) - 0.05\) over \(|Z_{\text{bal}}|\) and \(\rho\): red above nominal, blue below. The thick contour is the boundary \(z^*(\rho)\); open circles are the values in the table above. The dotted line marks \(|z| = 1\) (\(P_{\text{bal}} = 2\Phi(-1) \approx 0.3173\)) and the dashed line \(P_{\text{bal}} = 1/e\), which lies in the blue region for every \(\rho\). (b) Exact size against the second- and fourth-order expansion (D10.4) for \(\rho = 0.3\) and 0.5; all curves cross 0.05 near \(|z| = 1\).}
\end{figure}

\textbf{One-sided conditional size.} In the direction favoured by the imbalance (\(z > 0\), rejection for large \(T_u\)),

\[\alpha_1(z, \rho) = \Pr(T_u > q \mid z) = 1 - \Phi\!\left(\frac{q - \rho z}{\sqrt{1 - \rho^2}}\right). \qquad \text{(D10.5)}\]

\emph{Worked value.} At \(P_{\text{bal}} = 1/e\) and \(\rho = 0.5\): \(\rho z = 0.450\), \(s = 0.866\), \((1.960 - 0.450)/0.866 = 1.743\), \(\alpha_1 = 0.041\). At \(P_{\text{bal}} = 0.05\) and \(\rho = 0.9\): \(\rho z = 1.764\), \(s = 0.436\), \((1.960 - 1.764)/0.436 = 0.450\), \(\alpha_1 = 0.326\). The full set of values is Table A8. \(\alpha_1\) exceeds \(0.025\) whenever \(\rho z > q(1 - s)\), which is Senn's equation (4)~\cite{senn1989}; for large \(\rho\) and modest \(z\) the variance reduction dominates and \(\alpha_1\) falls below nominal. For a fixed \(0 < z < q\), \(\alpha_1\) is largest at \(\rho = z/q\), Senn's equation (5): at a balance \emph{P} value of \(1/e\) this gives \(\rho = 0.9005/1.960 = 0.459\) and \(\alpha_1 = 0.0409\), close to the \(\rho = 0.5\) entry of Table A8. \(\square\)

\begin{figure}
\centering
\includegraphics[width=0.66\textwidth,height=\textheight]{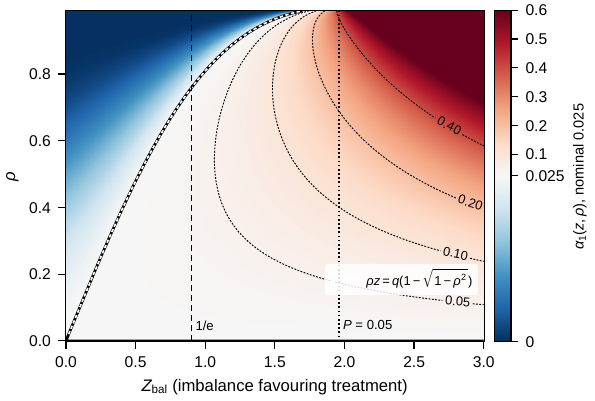}
\caption{One-sided conditional size in the direction favoured by the imbalance, \(\alpha_1(z,\rho)\) (D10.5), at nominal 0.025. The thick black contour, where \(\alpha_1 = 0.025\), coincides with the analytic curve \(\rho z = q(1 - \sqrt{1-\rho^2})\) (white dashes); to its right the error exceeds nominal. Dashed and dotted vertical lines mark balance \emph{P} values of \(1/e\) and 0.05.}
\end{figure}

\begin{table}[tbp]
\centering\small
\caption{Conditional one-sided type I error of the unadjusted test in the direction favoured by the imbalance, nominal 0.025}
\begin{tabular}{@{}llllll@{}}
\toprule\noalign{}
Balance \emph{P} & ρ = 0.1 & ρ = 0.3 & ρ = 0.5 & ρ = 0.7 & ρ = 0.9 \\
\midrule\noalign{}

0.50 & 0.029 & 0.033 & 0.030 & 0.019 & 0.001 \\
1/e ≈ 0.368 & 0.030 & 0.038 & 0.041 & 0.031 & 0.004 \\
0.20 & 0.033 & 0.049 & 0.064 & 0.068 & 0.032 \\
0.05 & 0.038 & 0.075 & 0.129 & 0.205 & 0.326 \\
0.01 & 0.044 & 0.107 & 0.219 & 0.413 & 0.794 \\
\bottomrule\noalign{}
\end{tabular}
\end{table}

\hypertarget{d11.-precision-reading-of-the-1e-threshold}{%
\subsection{D11. Precision reading of the 1/e threshold}\label{d11.-precision-reading-of-the-1e-threshold}}

Suppose complete-randomization allocations are retained only when one normal covariate's balance \emph{P} value is at least \(1/e\) (rerandomization with that criterion), that is \(|Z_{\text{bal}}| \le a\) with \(a = \Phi^{-1}(1 - 1/(2e)) = 0.9005\). Note that \(\Pr(|Z| \le a) = 1 - 1/e = 0.6321\) by construction.

\textbf{Truncated variance.} For \(Z \sim N(0, 1)\) truncated to \(|Z| \le a\), the mean is 0 by symmetry, and

\[\operatorname{Var}(Z \mid |Z| \le a) = 1 - \frac{2a\,\phi(a)}{2\Phi(a) - 1} = 1 - \frac{2(0.9005)(0.2662)}{0.6321} = 0.2422. \qquad \text{(D11.1)}\]

\textbf{Equality with Morgan and Rubin's \(v_a\).} For \(Q \sim \chi^2_k\) with density \(f_k\), the identity \(q f_k(q) = k f_{k+2}(q)\) gives \(\operatorname{E}\{Q\,\mathbf 1(Q \le c)\} = k\,\Pr(\chi^2_{k+2} \le c)\). With \(k = 1\), \(Q = Z^2\) and \(c = a^2\),

\[\operatorname{E}(Z^2 \mid Z^2 \le a^2) = \frac{\Pr(\chi^2_3 \le a^2)}{\Pr(\chi^2_1 \le a^2)} = v_a, \qquad \text{(D11.2)}\]

which is Morgan and Rubin's variance-reduction factor for rerandomization with one covariate and acceptance threshold \(a^2\)~\cite{morgan2012}. Numerically \(v_a = 0.2422\), a 75.8\% reduction in the variance of the covariate mean difference. By their result, the variance of \(\hat\tau\) falls by \(100(1 - v_a)R^2\%\), where \(R^2\) is the squared multiple correlation between the outcome and the covariates~\cite{morgan2012}. \(\square\)

\begin{figure}
\centering
\includegraphics[width=0.62\textwidth,height=\textheight]{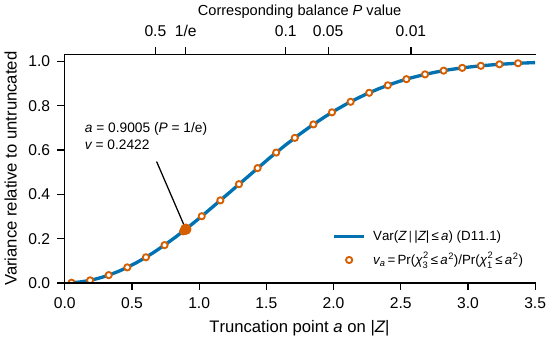}
\caption{The truncated-normal variance (D11.1, line) and Morgan and Rubin's \(v_a\) (D11.2, circles) are the same function of the truncation point \(a\). The filled point marks \(a = 0.9005\), the value at which the balance \emph{P} value equals \(1/e\), where both equal 0.2422. The upper axis gives the corresponding balance \emph{P} value.}
\end{figure}

\begin{figure}
\centering
\includegraphics[width=1\textwidth,height=\textheight]{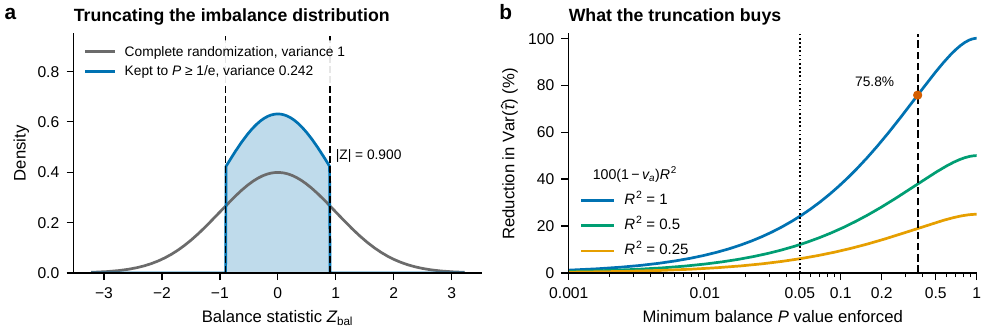}
\caption{A precision reading of the 1/e threshold. (a) Density of the balance statistic \(Z_{\text{bal}}\) under complete randomization (grey) and when restricted to \(|Z_{\text{bal}}| \le 0.900\), that is to balance \emph{P} values of at least \(1/e\) (blue): the variance falls from 1 to 0.242. (b) Reduction in the variance of the unadjusted treatment estimate if complete-randomization allocations were retained only when the balance \emph{P} value of one normal covariate is at least the value on the horizontal axis, \(100(1 - v_a)R^2\), for \(R^2 = 1\), 0.5 and 0.25; at \(1/e\) the reduction is \(75.8\% \times R^2\). Sequential schemes do not truncate the distribution this cleanly.}
\end{figure}

\hypertarget{d12.-non-collapsibility-an-illustration}{%
\subsection{D12. Non-collapsibility: an illustration}\label{d12.-non-collapsibility-an-illustration}}

Take a binary covariate with two equal strata, perfectly balanced between arms, control-arm risks \(0.10\) and \(0.60\), and a conditional odds ratio of \(8.0\) in both strata. Treated-arm risks are \(8 \times (0.1/0.9)/\{1 + 8 \times (0.1/0.9)\} = 0.4706\) and \(8 \times 1.5/(1 + 12) = 0.9231\). The marginal risks are \(p_C = 0.350\) and \(p_T = 0.6968\), giving a marginal odds ratio of

\[\frac{0.6968/0.3032}{0.350/0.650} = \frac{2.298}{0.5385} = 4.27. \qquad \text{(D12.1)}\]

The marginal odds ratio is closer to 1 than the common conditional odds ratio although the covariate is perfectly balanced. There is no bias in either estimand; they are different quantities~\cite{gail1984,fda2023}. On the risk-difference scale the stratum-specific differences (\(0.371\) and \(0.323\)) average exactly to the marginal difference (\(0.347\)), illustrating collapsibility of that scale. \(\square\)

\begin{table}[tbp]
\centering\small
\caption{Chance-imbalance bias versus non-collapsibility}
\begin{tabular}{@{}
  >{\raggedright\arraybackslash}p{(\columnwidth - 4\tabcolsep) * \real{0.3333}}
  >{\raggedright\arraybackslash}p{(\columnwidth - 4\tabcolsep) * \real{0.3333}}
  >{\raggedright\arraybackslash}p{(\columnwidth - 4\tabcolsep) * \real{0.3333}}@{}}
\toprule\noalign{}
\begin{minipage}[b]{\linewidth}\raggedright
Attribute
\end{minipage} & \begin{minipage}[b]{\linewidth}\raggedright
Chance-imbalance bias
\end{minipage} & \begin{minipage}[b]{\linewidth}\raggedright
Non-collapsibility
\end{minipage} \\
\midrule\noalign{}

Present with perfect balance? & No & Yes \\
Direction & Random: sign of βΔ & Towards the null for odds and hazard ratios \\
Scales affected & All, including mean and risk differences & Odds and hazard ratios; not mean or risk differences, risk ratios or log-linear models \\
Grows with & Imbalance × prognostic strength & Prognostic strength of omitted covariates \\
Nature & Conditional bias in this trial & Difference between conditional and marginal estimands \\
Narrowed by the randomization design? & Yes, in spread & No \\
Removed by adjustment? & Yes & Adjustment changes the estimand \\
\bottomrule\noalign{}
\end{tabular}
\end{table}

\begin{figure}
\centering
\includegraphics[width=1\textwidth,height=\textheight]{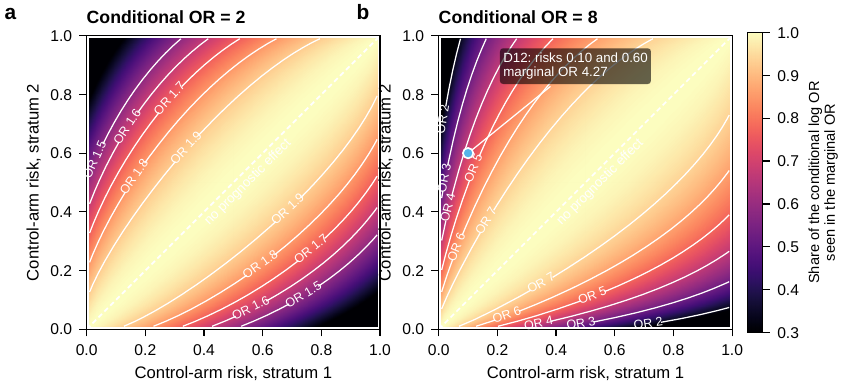}
\caption{Non-collapsibility of the odds ratio under perfect balance, over all pairs of stratum risks. Two equal strata, perfectly balanced between arms, with the control-arm event risks shown on the axes and a common conditional odds ratio of 2 (a) or 8 (b). Colour shows the share of the conditional log odds ratio retained by the marginal (unadjusted) odds ratio; white contours show the marginal odds ratio. On the diagonal the strata have equal risk, the covariate is not prognostic and there is no attenuation; attenuation grows with the risk difference between the strata although there is no imbalance. The point in (b) is the example above (risks 0.10 and 0.60, marginal odds ratio 4.27).}
\end{figure}

\begin{figure}
\centering
\includegraphics[width=1\textwidth,height=\textheight]{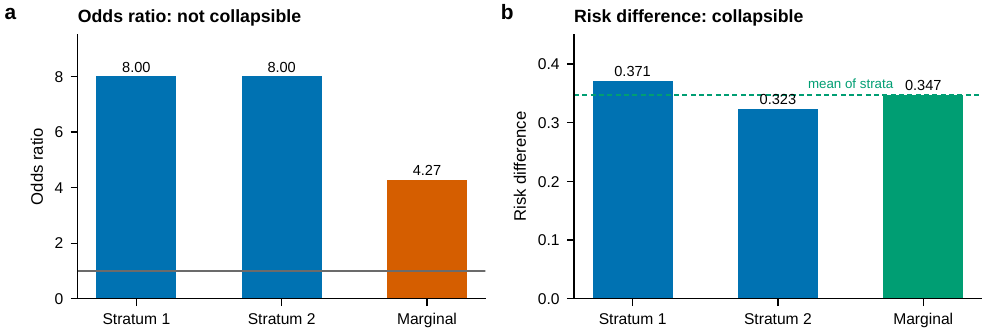}
\caption{The non-collapsibility example of D12: two equal, perfectly balanced strata with control-arm risks 0.10 and 0.60 and a conditional odds ratio of 8.0. (a) The marginal odds ratio, 4.27, is closer to 1 than the common conditional odds ratio. (b) On the risk-difference scale the stratum differences (0.371 and 0.323) average exactly to the marginal difference (0.347).}
\end{figure}

\hypertarget{d13.-separating-schemes-with-a-single-cut-off}{%
\subsection{D13. Separating schemes with a single cut-off}\label{d13.-separating-schemes-with-a-single-cut-off}}

\textbf{Criterion.} A score flags an allocation when its balance \emph{P} value is at most \(c\). Comparing a scheme with distribution function \(F\) for the balance \emph{P} value against complete randomization, under which \(P \sim U(0, 1)\), we measure the separation by the reduction in the probability of a flag,

\[D(c) = c - F(c). \qquad \text{(D13.1)}\]

Maximizing (D13.1) maximizes the vertical gap between the two distribution functions, a one-sided Kolmogorov or population Youden-type criterion with the two designs as the classes; at a differentiable interior maximum the two densities are equal, and each finite-improvement model below has a unique interior maximum. Other criteria, finite improvements or discrete tests can select other cut-offs.

\textbf{Beta shift.} Suppose the scheme shifts balance \emph{P} values upward within the Beta\((1 + \delta, 1)\) family, \(F(c) = c^{1+\delta}\) with \(\delta > 0\). Then

\[D(c) = c - c^{1 + \delta} = \delta\,(-c\ln c) + O(\delta^2), \qquad \frac{d}{dc}(-c\ln c) = -1 - \ln c, \qquad \frac{d^2}{dc^2}(-c\ln c) = -\frac{1}{c}, \qquad \text{(D13.2)}\]

so the first-order separation is maximized uniquely at \(c = 1/e\). At a finite improvement the densities cross where \((1 + \delta)c^\delta = 1\), that is at \(c_\delta = (1 + \delta)^{-1/\delta}\), which decreases to \(1/e\) as \(\delta \to 0\): 0.3697 at \(\delta = 0.01\), 0.3855 at \(\delta = 0.1\), 0.4444 at \(\delta = 0.5\) and 0.5 at \(\delta = 1\). The corresponding largest gap, \(c_\delta - c_\delta^{1+\delta}\), is the maximum vertical distance between a distribution function and its Lehmann alternative evaluated by Wellek~\cite{wellek2021} (his equations 2 and 4, with \(\theta = 1 + \delta\)); the limit \(1/e\) as \(\delta \to 0\) is the local result used here.

\emph{Relation to the Sellke--Bayarri--Berger bound.} For alternatives Beta\((\xi, 1)\) with \(0 < \xi \le 1\), which favour small \emph{P} values, the largest likelihood ratio against the uniform density is \(\sup_\xi \xi p^{\xi - 1}\), attained at \(\xi = -1/\ln p\) for \(p < 1/e\), which gives the bound \(B(p) = -e\,p\ln p\) of D7~\cite{sellke2001,heldott2018}. A better-balancing scheme corresponds to \(\xi = 1 + \delta > 1\) and favours large \emph{P} values. The two calculations share the function \(-p\ln p\) and its maximum at \(1/e\), not an interpretation.

\textbf{Normal contraction.} Suppose instead that the scheme contracts the balance statistic, \(Z_{\text{bal}} \sim N(0, s^2)\) with \(s = 1 - \epsilon\). With \(a = \Phi^{-1}(1 - c/2)\),

\[D(c) = 2\{\Phi(-a) - \Phi(-a/s)\} = 2\epsilon\,a\,\phi(a) + O(\epsilon^2), \qquad \frac{d}{da}\{a\phi(a)\} = (1 - a^2)\,\phi(a), \qquad \text{(D13.3)}\]

so the first-order separation is maximized uniquely at \(a = 1\), that is at \(c = 2\Phi(-1) = 0.3173\ldots\), the sharp two-sided boundary of D10. At a finite contraction the densities cross at \(a_s = s\sqrt{2\ln(1/s)/(1 - s^2)}\), which decreases from 1 as the contraction grows; the optimal cut-off rises: 0.3197 at \(\epsilon = 0.01\), 0.3432 at 0.1, 0.4077 at 0.3 and 0.4966 at 0.5.

\textbf{Efficiency.} Dividing each local separation by its maximum under the same model:

\begin{center}\small
\begin{tabular}{@{}
  >{\raggedright\arraybackslash}p{(\columnwidth - 16\tabcolsep) * \real{0.1111}}
  >{\raggedright\arraybackslash}p{(\columnwidth - 16\tabcolsep) * \real{0.1111}}
  >{\raggedright\arraybackslash}p{(\columnwidth - 16\tabcolsep) * \real{0.1111}}
  >{\raggedright\arraybackslash}p{(\columnwidth - 16\tabcolsep) * \real{0.1111}}
  >{\raggedright\arraybackslash}p{(\columnwidth - 16\tabcolsep) * \real{0.1111}}
  >{\raggedright\arraybackslash}p{(\columnwidth - 16\tabcolsep) * \real{0.1111}}
  >{\raggedright\arraybackslash}p{(\columnwidth - 16\tabcolsep) * \real{0.1111}}
  >{\raggedright\arraybackslash}p{(\columnwidth - 16\tabcolsep) * \real{0.1111}}
  >{\raggedright\arraybackslash}p{(\columnwidth - 16\tabcolsep) * \real{0.1111}}@{}}
\toprule\noalign{}
\begin{minipage}[b]{\linewidth}\raggedright
Cut-off \(c\)
\end{minipage} & \begin{minipage}[b]{\linewidth}\raggedright
0.25
\end{minipage} & \begin{minipage}[b]{\linewidth}\raggedright
0.30
\end{minipage} & \begin{minipage}[b]{\linewidth}\raggedright
\(2\Phi(-1)\)
\end{minipage} & \begin{minipage}[b]{\linewidth}\raggedright
0.35
\end{minipage} & \begin{minipage}[b]{\linewidth}\raggedright
\(1/e\)
\end{minipage} & \begin{minipage}[b]{\linewidth}\raggedright
0.40
\end{minipage} & \begin{minipage}[b]{\linewidth}\raggedright
0.45
\end{minipage} & \begin{minipage}[b]{\linewidth}\raggedright
0.50
\end{minipage} \\
\midrule\noalign{}
Beta shift & 0.942 & 0.982 & 0.990 & 0.999 & 1 & 0.996 & 0.977 & 0.942 \\
Normal contraction & 0.979 & 0.999 & 1 & 0.996 & 0.990 & 0.974 & 0.936 & 0.886 \\
\bottomrule\noalign{}
\end{tabular}
\end{center}

At \(1/e\) the normal-contraction efficiency is 0.98979, and at \(2\Phi(-1)\) the Beta-shift efficiency is 0.99009: each cut-off retains about 99\% of the other model's best local separation. Every cut-off from 0.30 to 0.40 retains at least 0.97 under both models. Together with D7 and D10, three criteria (the Bayes-factor bound, uniform two-sided error control and local separation of schemes) point to the band \([2\Phi(-1), 1/e]\) for a single balance \emph{P} value.

\textbf{Worst of \(k\) covariates.} For \(k\) independent balance \emph{P} values, each improved in the same way, a flag is \(\min_j P_j \le c\), with probability \(1 - (1 - c)^k\) under complete randomization (D8). The first-order separation acquires the factor \(k(1 - c)^{k-1}\): \(\delta\,k(1 - c)^{k-1}(-c\ln c)\) under the Beta shift and \(2\epsilon\,k(1 - c)^{k-1}a\phi(a)\) under normal contraction. For one improved covariate to change whether any flag fires, all the others must clear the cut-off, so for \(k > 1\) the best cut-off falls with \(k\):

\begin{center}\small
\begin{tabular}{@{}
  >{\raggedright\arraybackslash}p{(\columnwidth - 8\tabcolsep) * \real{0.2000}}
  >{\raggedright\arraybackslash}p{(\columnwidth - 8\tabcolsep) * \real{0.2000}}
  >{\raggedright\arraybackslash}p{(\columnwidth - 8\tabcolsep) * \real{0.2000}}
  >{\raggedright\arraybackslash}p{(\columnwidth - 8\tabcolsep) * \real{0.2000}}
  >{\raggedright\arraybackslash}p{(\columnwidth - 8\tabcolsep) * \real{0.2000}}@{}}
\toprule\noalign{}
\begin{minipage}[b]{\linewidth}\raggedright
\(k\)
\end{minipage} & \begin{minipage}[b]{\linewidth}\raggedright
Best \(c\), Beta shift
\end{minipage} & \begin{minipage}[b]{\linewidth}\raggedright
Best \(c\), normal contraction
\end{minipage} & \begin{minipage}[b]{\linewidth}\raggedright
Efficiency of \(1/e\), Beta
\end{minipage} & \begin{minipage}[b]{\linewidth}\raggedright
Efficiency of \(1/e\), normal
\end{minipage} \\
\midrule\noalign{}
1 & 0.3679 & 0.3173 & 1.000 & 0.990 \\
2 & 0.2356 & 0.2093 & 0.893 & 0.841 \\
5 & 0.1176 & 0.1081 & 0.385 & 0.343 \\
10 & 0.0657 & 0.0616 & 0.061 & 0.053 \\
\bottomrule\noalign{}
\end{tabular}
\end{center}

For a worst-of-\(k\) score, \(1/e\) is therefore a fixed per-covariate reference, read against the complete-randomization benchmark for the same covariates, not the cut-off that best separates schemes. With correlated or discrete covariates the benchmark should be taken from simulation.

\textbf{Several prognostic directions.} Let \(Z \sim N_k(0, R)\) be the standardized treatment-minus-control imbalances under complete randomization, with \(R\) positive definite, and let prognostic weights \(w\) be fixed in these coordinates. Fix \(w \ne 0\) and put \(U = w^\top Z/\sqrt{w^\top R w}\), with observed value \(u = w^\top z/\sqrt{w^\top R w}\). Assume that under the null the outcome statistic is \(T = rU + \sqrt{1 - r^2}\,E\), where \(E\) is standard normal and independent of the whole vector \(Z\). Conditional on \(Z = z\), (D10.3) then applies with \(u\) in place of \(z\) and \(\rho = r\).

\emph{Example.} Take \(R = I_2\), \(w = (1, 1)\) and \(r = 0.5\). The allocations \(z = (0.88, 0.88)\) and \(z = (0.88, -0.88)\) both have balance \emph{P} values \(0.3789\) for each covariate, above \(1/e\). The first gives \(u = 1.2445\) and a two-sided conditional size of 0.0626; the second gives \(u = 0\) and 0.0236. No rule that uses only the two unsigned \emph{P} values can distinguish them, and a worst-\emph{P} value above \(1/e\) does not by itself bound the conditional error when several covariates are prognostic.

\emph{Uniform guarantee.} Maximizing \(u^2\) over \(w\) gives \(z^\top R^{-1} z\) by the Cauchy--Schwarz inequality, with equality at \(w \propto R^{-1}z\). Hence, by D10, \(z^\top R^{-1} z \le 1\) is necessary and sufficient, in this model, for two-sided conditional error control uniform in prognostic direction and strength. This is a Mahalanobis criterion of the kind used in rerandomization~\cite{morgan2012}. For a common marginal limit \(|z_j| \le a\) the largest \(a\) for which every point of the box satisfies it is \((\max_{v \in \{-1, 1\}^k} v^\top R^{-1} v)^{-1/2}\); with \(R = I\) this is \(1/\sqrt{k}\), a balance \emph{P} value of \(2\Phi(-1/\sqrt{k})\): 0.4795 for \(k = 2\) and 0.6547 for \(k = 5\) (0.7518 for two covariates with correlation 0.8). These illustrate the cost of protecting every direction and are not proposed as thresholds.

\emph{Checks.} Every value above is recomputed by imbalance\_calculations.py (Additional file 2), which also confirms each optimum by numerical maximization.

\hypertarget{m.-monte-carlo-checks}{%
\subsection{M. Monte Carlo checks}\label{m.-monte-carlo-checks}}

The script in Additional file 2 checks the analytic results by simulation (default settings; seeds fixed in the code).

\begin{center}\small
\begin{tabular}{@{}
  >{\raggedright\arraybackslash}p{(\columnwidth - 6\tabcolsep) * \real{0.2500}}
  >{\raggedright\arraybackslash}p{(\columnwidth - 6\tabcolsep) * \real{0.2500}}
  >{\raggedright\arraybackslash}p{(\columnwidth - 6\tabcolsep) * \real{0.2500}}
  >{\raggedright\arraybackslash}p{(\columnwidth - 6\tabcolsep) * \real{0.2500}}@{}}
\toprule\noalign{}
\begin{minipage}[b]{\linewidth}\raggedright
Check
\end{minipage} & \begin{minipage}[b]{\linewidth}\raggedright
Setting
\end{minipage} & \begin{minipage}[b]{\linewidth}\raggedright
Simulated
\end{minipage} & \begin{minipage}[b]{\linewidth}\raggedright
Analytic
\end{minipage} \\
\midrule\noalign{}
Slope of final-value \(\hat\tau\) on \(\Delta_X\) (D1) & \(n = 100\)/arm, \(\rho = \beta = 0.5\), 20 000 trials & 0.496 & 0.50 \\
Slope of change-score estimate on \(\Delta_X\) (D3) & same & −0.504 & −0.50 \\
Slope of ANCOVA estimate on \(\Delta_X\) (D3) & same & −0.004 & 0 \\
SD ratio, ANCOVA / final value (D3.4) & same & 0.869 & 0.866 \\
Two-sided conditional size, \(Z_{\text{bal}} \in [0.85, 0.95]\) (D10) & \(\rho = 0.5\), \(10^7\) draws & 0.0436 & 0.0433 \\
One-sided conditional size, same bin (D10) & same & 0.0409 & 0.0406 \\
Slope of crude risk difference on share gap (D4) & \(n = 500\)/arm, \(r_H - r_L = 0.389\), 200 000 trials & 0.387 & 0.389 \\
Mean ANCOVA variance inflation factor (D3.5) & normal covariate, \(n = 20\)/arm, 200 000 trials & 1.0276 & 1.0278 \\
\bottomrule\noalign{}
\end{tabular}
\end{center}

The analytic values in the D10 rows are averaged over the bin with weights proportional to \(\phi(z)\). The simulated values were recorded with Python 3.13 and NumPy 2.5; other software versions can change the last digit of a Monte Carlo value (for example, 0.388 for the risk-difference slope).

\textbf{Small trials.} The D10 results treat variances as known. To check them in small trials, Additional file 2 simulates normal data with estimated variances: the balance test and the outcome test are pooled-variance two-sample t-tests with \(2n - 2\) degrees of freedom, and the outcome test rejects at the t quantile. Conditional errors are estimated in bins of the balance \emph{P} value and compared with the normal-theory values of (D10.3) and (D10.5) averaged over the same bin (1 000 000 simulated trials per row).

\begin{center}\small
\begin{tabular}{@{}
  >{\raggedright\arraybackslash}p{(\columnwidth - 12\tabcolsep) * \real{0.1429}}
  >{\raggedright\arraybackslash}p{(\columnwidth - 12\tabcolsep) * \real{0.1429}}
  >{\raggedright\arraybackslash}p{(\columnwidth - 12\tabcolsep) * \real{0.1429}}
  >{\raggedright\arraybackslash}p{(\columnwidth - 12\tabcolsep) * \real{0.1429}}
  >{\raggedright\arraybackslash}p{(\columnwidth - 12\tabcolsep) * \real{0.1429}}
  >{\raggedright\arraybackslash}p{(\columnwidth - 12\tabcolsep) * \real{0.1429}}
  >{\raggedright\arraybackslash}p{(\columnwidth - 12\tabcolsep) * \real{0.1429}}@{}}
\toprule\noalign{}
\begin{minipage}[b]{\linewidth}\raggedright
\(n\) per arm
\end{minipage} & \begin{minipage}[b]{\linewidth}\raggedright
\(\rho\)
\end{minipage} & \begin{minipage}[b]{\linewidth}\raggedright
Balance P value
\end{minipage} & \begin{minipage}[b]{\linewidth}\raggedright
Two-sided, simulated
\end{minipage} & \begin{minipage}[b]{\linewidth}\raggedright
Two-sided, normal theory
\end{minipage} & \begin{minipage}[b]{\linewidth}\raggedright
One-sided, simulated
\end{minipage} & \begin{minipage}[b]{\linewidth}\raggedright
One-sided, normal theory
\end{minipage} \\
\midrule\noalign{}
10 & 0.5 & 0.33--0.41 & 0.0441 & 0.0432 & 0.0408 & 0.0405 \\
10 & 0.5 & 0.04--0.06 & 0.1233 & 0.1296 & 0.1224 & 0.1293 \\
10 & 0.9 & 0.33--0.41 & 0.0061 & 0.0042 & 0.0061 & 0.0042 \\
10 & 0.9 & 0.04--0.06 & 0.3262 & 0.3291 & 0.3262 & 0.3291 \\
25 & 0.5 & 0.33--0.41 & 0.0439 & 0.0432 & 0.0407 & 0.0405 \\
25 & 0.5 & 0.04--0.06 & 0.1285 & 0.1296 & 0.1278 & 0.1293 \\
25 & 0.9 & 0.33--0.41 & 0.0046 & 0.0042 & 0.0046 & 0.0042 \\
25 & 0.9 & 0.04--0.06 & 0.3277 & 0.3291 & 0.3277 & 0.3291 \\
\bottomrule\noalign{}
\end{tabular}
\end{center}

The largest difference is 0.6 percentage points, at 10 patients per arm and a balance \emph{P} value near 0.05, where the t-test is slightly conservative relative to the normal approximation; at 25 per arm no difference exceeds 0.2 points. The normal-theory results are therefore adequate for the trial sizes at which design-stage scoring is used.

\begin{figure}
\centering
\includegraphics[width=1\textwidth,height=\textheight]{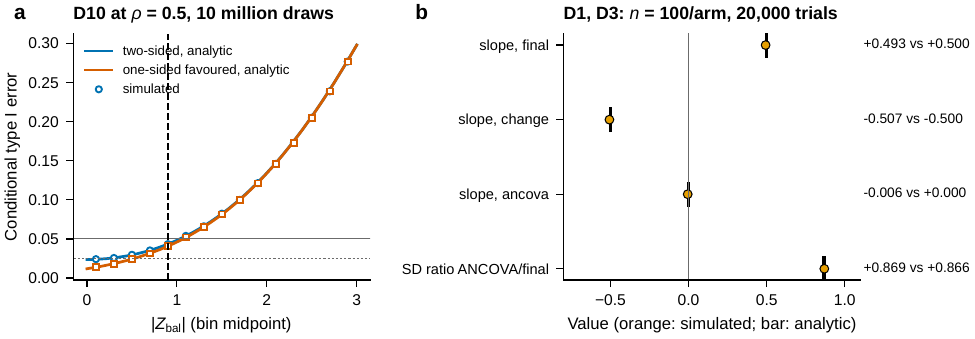}
\caption{Monte Carlo checks drawn by Additional file 3. (a) Conditional two-sided (circles) and one-sided (squares) type I error of the unadjusted test in bins of \(|Z_{\text{bal}}|\), from 10 million simulated \((Z_{\text{bal}}, T_u)\) pairs at \(\rho = 0.5\), against the analytic curves of D10.3 and D10.5. (b) Slopes of the final-value, change-score and ANCOVA estimates on \(\Delta_X\), and the ratio of their standard deviations, from 20 000 simulated trials with 100 patients per arm and \(\rho = 0.5\) (orange), against the analytic values (black bars). Seeds differ from Additional file 2, so the third decimal differs from the table above.}
\end{figure}

\hypertarget{n.-simulated-comparison-of-randomization-schemes}{%
\subsection{N. Simulated comparison of randomization schemes}\label{n.-simulated-comparison-of-randomization-schemes}}

Additional file 4 simulates allocations of the same patients under six schemes and scores each allocation by its worst balance \emph{P} value over five covariates, using pooled-variance two-sample t tests (also for the two binary covariates). Covariates: three continuous and two binary, with prevalences 0.30 and 0.50, generated by thresholding an exchangeably correlated latent normal vector (correlation 0.2). Outcome: \(Y = \rho g + \sqrt{1-\rho^2}\,\varepsilon\), where \(g\) is a standardized prognostic index with weights 0.60, 0.30, 0.15, 0.50 and 0.25 on the five covariates, and there is no treatment effect. Schemes: complete randomization with equal arm sizes; permuted blocks of 4; Efron's biased coin with bias 2/3; permuted blocks of 4 within the four strata formed by the binary covariates; Pocock--Simon minimization on the two binary covariates and the first continuous covariate split at 0, its pre-specified population median, so that every factor level is known at enrolment, assigning the preferred arm with probability 0.80; and, as a reference, rerandomization on the Mahalanobis distance of the five balance statistics, accepting an allocation within the 5\% quantile of the chi-squared distribution with the covariate correlation pooled over the simulated cohorts. Sample sizes are targets of \(n\) patients per arm (\(2n\) in total); final arm sizes can differ under Efron's coin, stratified blocks and minimization. Balance t statistics keep \(\pm\infty\) for perfect separation of a binary covariate (balance \emph{P} = 0); none occurred. Predictability is the probability of guessing the next assignment correctly by choosing the arm that is behind overall (complete randomization, blocks, biased coin, rerandomization), behind within the patient's stratum (stratified blocks), or preferred by the minimization rule (minimization). For rerandomization this count-based guesser is not the best prediction open to someone who knows the covariates and the acceptance rule. Version 1.1 of Additional file 4 split the first covariate at the median of the whole simulated cohort, which is not known when early patients enrol; with the prospective cut-off the minimization score at 50 per arm changed from 0.6583 to 0.6561 (paired Monte Carlo SE 0.0036), and no other scheme changed. Seed 20261122; 20 000 allocations per scheme and sample size.

\begin{table}[tbp]
\centering\small
\caption{Worst-of-k benchmark under complete randomization, simulated against equation (7)}
\begin{tabular}{@{}
  >{\raggedright\arraybackslash}p{(\columnwidth - 8\tabcolsep) * \real{0.4400}}
  >{\raggedright\arraybackslash}p{(\columnwidth - 8\tabcolsep) * \real{0.1000}}
  >{\raggedright\arraybackslash}p{(\columnwidth - 8\tabcolsep) * \real{0.1200}}
  >{\raggedright\arraybackslash}p{(\columnwidth - 8\tabcolsep) * \real{0.2000}}
  >{\raggedright\arraybackslash}p{(\columnwidth - 8\tabcolsep) * \real{0.1400}}@{}}
\toprule\noalign{}
\begin{minipage}[b]{\linewidth}\raggedright
Covariates scored
\end{minipage} & \begin{minipage}[b]{\linewidth}\raggedright
\(n\)/arm
\end{minipage} & \begin{minipage}[b]{\linewidth}\raggedright
Allocations
\end{minipage} & \begin{minipage}[b]{\linewidth}\raggedright
Simulated Pr(worst \emph{P} ≤ 1/e)
\end{minipage} & \begin{minipage}[b]{\linewidth}\raggedright
Equation (7)
\end{minipage} \\
\midrule\noalign{}

1 independent continuous & 100 & 100 000 & 0.3668 & 0.3679 \\
2 independent continuous & 100 & 100 000 & 0.5997 & 0.6004 \\
3 independent continuous & 100 & 100 000 & 0.7473 & 0.7474 \\
5 independent continuous & 100 & 100 000 & 0.8990 & 0.8991 \\
10 independent continuous & 100 & 100 000 & 0.9901 & 0.9898 \\
5 independent continuous & 10 & 20 000 & 0.8969 & 0.8991 \\
5 independent continuous & 25 & 20 000 & 0.8969 & 0.8991 \\
5 independent continuous & 50 & 20 000 & 0.8990 & 0.8991 \\
5 independent continuous & 100 & 20 000 & 0.9015 & 0.8991 \\
5 independent continuous & 500 & 20 000 & 0.8995 & 0.8991 \\
5 independent continuous & 100 & 100 000 & 0.8980 & 0.8991 \\
5 continuous, correlation 0.2 & 100 & 100 000 & 0.8898 & 0.8991 \\
5 continuous, correlation 0.5 & 100 & 100 000 & 0.8357 & 0.8991 \\
3 continuous, 2 binary (0.30, 0.50), correlation 0.2 & 100 & 100 000 & 0.8916 & 0.8991 \\
3 continuous, 2 binary (0.10, 0.50), correlation 0.2 & 100 & 100 000 & 0.8885 & 0.8991 \\
\bottomrule\noalign{}
\end{tabular}
\end{table}

Monte Carlo standard errors: at most 0.0015 with 100 000 allocations and about 0.002 with 20 000. Correlation is the exchangeable correlation of the latent normal vector; binary covariates are thresholded at the stated prevalences. Correlation among covariates lowers the benchmark, so a scheme's score is read against a benchmark simulated for the covariates actually scored.

\begin{table}[tbp]
\centering\small
\caption{Six schemes compared: balance score, predictability and precision}
\begin{tabular}{@{}
  >{\raggedright\arraybackslash}p{(\columnwidth - 10\tabcolsep) * \real{0.1000}}
  >{\raggedright\arraybackslash}p{(\columnwidth - 10\tabcolsep) * \real{0.3500}}
  >{\raggedright\arraybackslash}p{(\columnwidth - 10\tabcolsep) * \real{0.1250}}
  >{\raggedright\arraybackslash}p{(\columnwidth - 10\tabcolsep) * \real{0.1750}}
  >{\raggedright\arraybackslash}p{(\columnwidth - 10\tabcolsep) * \real{0.1250}}
  >{\raggedright\arraybackslash}p{(\columnwidth - 10\tabcolsep) * \real{0.1250}}@{}}
\toprule\noalign{}
\begin{minipage}[b]{\linewidth}\raggedright
Target \(n\)/arm
\end{minipage} & \begin{minipage}[b]{\linewidth}\raggedright
Scheme
\end{minipage} & \begin{minipage}[b]{\linewidth}\raggedright
Score
\end{minipage} & \begin{minipage}[b]{\linewidth}\raggedright
Pr(worst \emph{P} \textless{} 0.05)
\end{minipage} & \begin{minipage}[b]{\linewidth}\raggedright
Correct guess
\end{minipage} & \begin{minipage}[b]{\linewidth}\raggedright
ANCOVA SE ratio
\end{minipage} \\
\midrule\noalign{}

25 & Complete randomization & 0.8910 & 0.2364 & 0.579 & 1.000 \\
25 & Permuted blocks & 0.8903 & 0.2366 & 0.703 & 1.000 \\
25 & Efron coin & 0.8868 & 0.2338 & 0.621 & 1.000 \\
25 & Stratified blocks & 0.7326 & 0.1265 & 0.693 & 0.981 \\
25 & Minimization & 0.6826 & 0.0901 & 0.738 & 0.977 \\
25 & Rerandomization* & 0.0333 & 0.0000 & 0.579 & 0.954 \\
50 & Complete randomization & 0.8893 & 0.2272 & 0.558 & 1.000 \\
50 & Permuted blocks & 0.8871 & 0.2195 & 0.708 & 1.000 \\
50 & Efron coin & 0.8950 & 0.2303 & 0.623 & 1.000 \\
50 & Stratified blocks & 0.7212 & 0.1262 & 0.701 & 0.990 \\
50 & Minimization & 0.6561 & 0.0887 & 0.740 & 0.988 \\
50 & Rerandomization* & 0.0314 & 0.0000 & 0.558 & 0.978 \\
100 & Complete randomization & 0.8951 & 0.2260 & 0.542 & 1.000 \\
100 & Permuted blocks & 0.8961 & 0.2301 & 0.708 & 1.000 \\
100 & Efron coin & 0.8901 & 0.2228 & 0.624 & 1.000 \\
100 & Stratified blocks & 0.7307 & 0.1292 & 0.704 & 0.995 \\
100 & Minimization & 0.6411 & 0.0849 & 0.741 & 0.994 \\
100 & Rerandomization* & 0.0395 & 0.0000 & 0.542 & 0.989 \\
\bottomrule\noalign{}
\end{tabular}
\end{table}

Score: Pr(worst of five balance \emph{P} values ≤ 1/e). Correct guess: the guess rules of Section N. ANCOVA SE ratio: mean ANCOVA standard error relative to complete randomization on the same patients; the same for ρ = 0.3, 0.5 and 0.7 to three decimals. *Reference only: rerandomization needs every patient's covariates before allocation.

\begin{table}[tbp]
\centering\small
\caption{Six schemes compared: type I error of the unadjusted test and of ANCOVA}
\begin{tabular}{@{}
  >{\raggedright\arraybackslash}p{(\columnwidth - 16\tabcolsep) * \real{0.0833}}
  >{\raggedright\arraybackslash}p{(\columnwidth - 16\tabcolsep) * \real{0.2917}}
  >{\raggedright\arraybackslash}p{(\columnwidth - 16\tabcolsep) * \real{0.0833}}
  >{\raggedright\arraybackslash}p{(\columnwidth - 16\tabcolsep) * \real{0.0833}}
  >{\raggedright\arraybackslash}p{(\columnwidth - 16\tabcolsep) * \real{0.0833}}
  >{\raggedright\arraybackslash}p{(\columnwidth - 16\tabcolsep) * \real{0.1042}}
  >{\raggedright\arraybackslash}p{(\columnwidth - 16\tabcolsep) * \real{0.0833}}
  >{\raggedright\arraybackslash}p{(\columnwidth - 16\tabcolsep) * \real{0.0833}}
  >{\raggedright\arraybackslash}p{(\columnwidth - 16\tabcolsep) * \real{0.1042}}@{}}
\toprule\noalign{}
\begin{minipage}[b]{\linewidth}\raggedright
Target \(n\)/arm
\end{minipage} & \begin{minipage}[b]{\linewidth}\raggedright
Scheme
\end{minipage} & \begin{minipage}[b]{\linewidth}\raggedright
Two-sided, ρ = 0.3
\end{minipage} & \begin{minipage}[b]{\linewidth}\raggedright
ρ = 0.5
\end{minipage} & \begin{minipage}[b]{\linewidth}\raggedright
ρ = 0.7
\end{minipage} & \begin{minipage}[b]{\linewidth}\raggedright
Favoured direction, ρ = 0.3
\end{minipage} & \begin{minipage}[b]{\linewidth}\raggedright
ρ = 0.5
\end{minipage} & \begin{minipage}[b]{\linewidth}\raggedright
ρ = 0.7
\end{minipage} & \begin{minipage}[b]{\linewidth}\raggedright
ANCOVA, two-sided
\end{minipage} \\
\midrule\noalign{}

25 & Complete randomization & 0.0478 & 0.0497 & 0.0493 & 0.0376 & 0.0447 & 0.0487 & 0.0461 \\
25 & Permuted blocks & 0.0515 & 0.0515 & 0.0521 & 0.0416 & 0.0479 & 0.0515 & 0.0503 \\
25 & Efron coin & 0.0502 & 0.0499 & 0.0511 & 0.0382 & 0.0454 & 0.0505 & 0.0509 \\
25 & Stratified blocks & 0.0464 & 0.0417 & 0.0345 & 0.0337 & 0.0372 & 0.0338 & 0.0483 \\
25 & Minimization & 0.0457 & 0.0355 & 0.0215 & 0.0307 & 0.0290 & 0.0204 & 0.0505 \\
25 & Rerandomization* & 0.0397 & 0.0262 & 0.0101 & 0.0248 & 0.0190 & 0.0088 & 0.0491 \\
50 & Complete randomization & 0.0512 & 0.0515 & 0.0515 & 0.0393 & 0.0464 & 0.0507 & 0.0505 \\
50 & Permuted blocks & 0.0471 & 0.0467 & 0.0462 & 0.0352 & 0.0419 & 0.0456 & 0.0480 \\
50 & Efron coin & 0.0466 & 0.0461 & 0.0471 & 0.0355 & 0.0418 & 0.0466 & 0.0483 \\
50 & Stratified blocks & 0.0467 & 0.0410 & 0.0337 & 0.0341 & 0.0362 & 0.0328 & 0.0504 \\
50 & Minimization & 0.0421 & 0.0336 & 0.0197 & 0.0289 & 0.0270 & 0.0187 & 0.0485 \\
50 & Rerandomization* & 0.0411 & 0.0275 & 0.0112 & 0.0261 & 0.0204 & 0.0097 & 0.0503 \\
100 & Complete randomization & 0.0522 & 0.0515 & 0.0526 & 0.0411 & 0.0471 & 0.0520 & 0.0520 \\
100 & Permuted blocks & 0.0531 & 0.0544 & 0.0546 & 0.0420 & 0.0507 & 0.0544 & 0.0508 \\
100 & Efron coin & 0.0483 & 0.0501 & 0.0517 & 0.0384 & 0.0459 & 0.0512 & 0.0488 \\
100 & Stratified blocks & 0.0477 & 0.0419 & 0.0351 & 0.0357 & 0.0373 & 0.0342 & 0.0520 \\
100 & Minimization & 0.0441 & 0.0325 & 0.0175 & 0.0286 & 0.0256 & 0.0162 & 0.0505 \\
100 & Rerandomization* & 0.0425 & 0.0277 & 0.0123 & 0.0276 & 0.0214 & 0.0109 & 0.0507 \\
\bottomrule\noalign{}
\end{tabular}
\end{table}

Unadjusted: two-sample t test, nominal 0.05 two-sided; favoured direction: rejections in the direction favoured by the imbalance of the prognostic index (0.025 if the imbalance were unrelated to the outcome). ANCOVA: pre-specified adjustment for all five covariates; its rate is the same for every ρ in this design, because the covariates span the prognostic index and the residual draws are only rescaled, so the three ρ settings are not separate checks. With 20 000 allocations the Monte Carlo standard error of an error rate near 0.05 is about 0.0015. Values are shown to four decimals so that every number quoted in the text can be read from the table. *Reference only.

\end{document}